\documentclass[compsoc]{IEEEtran}

\usepackage[english]{babel}
\usepackage[utf8]{inputenc}

\usepackage{verbatim}
 \usepackage[cmex10]{amsmath}
\ifCLASSOPTIONcompsoc
\usepackage[caption=false,font=normalsize,labelfont=sf,textfont=sf]{subfig}
\else
\usepackage[caption=false,font=footnotesize]{subfig}
\fi
\usepackage{cancel}
\ifCLASSOPTIONcompsoc
\usepackage[nocompress]{cite}
\else
\usepackage{cite}
\fi
\usepackage{mdwmath}
\usepackage{mdwtab}
\usepackage{placeins}
\usepackage{graphicx}

\newtheorem{lemma}{Lemma}[section]

\begin{document}
\title{Service Provisioning in Mobile Environments through Opportunistic Computing}


\author{\IEEEauthorblockN{
Davide Mascitti\IEEEauthorrefmark{1},
Marco Conti\IEEEauthorrefmark{1},
Andrea Passarella\IEEEauthorrefmark{1},
Laura Ricci\IEEEauthorrefmark{2},
Sajal K. Das\IEEEauthorrefmark{3}}
%\IEEEauthorblockA

{\IEEEauthorrefmark{1}
IIT-CNR,
Via G. Moruzzi 1
56124, Pisa, Italy\\
\IEEEauthorrefmark{2} Department of Computer Science, University of Pisa, Italy \\
\IEEEauthorrefmark{3} Department of Computer Science, Missouri University of Science and Technology, MO 65409, USA}

email: \{davide.mascitti, marco.conti, a.passarella\}@iit.cnr.it, 
laura.ricci@unipi.it, sdas@mst.edu}

\IEEEoverridecommandlockouts
%\IEEEpubid{\makebox[\columnwidth]{978-1-4673-2480-9/13/\$31.00 ~\copyright~2013 IEEE \hfill} \hspace{\columnsep}\makebox[\columnwidth]{ }}

\IEEEtitleabstractindextext{%
\begin{abstract}
Opportunistic computing is a paradigm for completely self-organised pervasive networks. Instead of relying only on fixed infrastructures as the cloud, users’ devices act as service providers for each other. They use pairwise contacts to collect information about services provided and amount of time to provide them by the encountered nodes. At each node, upon generation of a service request, this information is used to choose the most efficient service, or composition of services, that satisfy that request, based on local knowledge. Opportunistic computing can be exploited in several scenarios, including mobile social networks, IoT and Internet 4.0. In this paper we propose an opportunistic computing algorithm based on an analytical model, which ranks the available (composition of) services, based on their expected completion time. Through the model, a service requesters picks the one that is expected to be the best. Experiments show that the algorithm is accurate in ranking services, thus providing an effective service-selection policy. Such a policy achieves significantly lower service provisioning times compared to other reference policies. Its performance is tested in a wide range of scenarios varying the nodes mobility, the size of input/output parameters, the level of resource congestion, the computational complexity of service executions.
\end{abstract}

}

\maketitle

\IEEEraisesectionheading{
\section{Introduction}\label{sec:Introduction}}

\IEEEPARstart{P}{ervasive} personal users' devices (e.g., smartphones, tablets, etc.) and IoT devices (e.g., sensors)  in the physical environment are one of the key factors pushing the BigData era~\cite{cisco2016,Vincentelli:2015aa}. The number of such devices and the amount of data they generate are expected to grow exponentially over the next few years \cite{cisco2016}. To elaborate this huge amount of data, cloud platforms are typically used, with mobile devices acting as mere data generators and consumers of the output of data elaboration. This approach might not be a solution working well in all scenarios. Even next generation cellular technologies might not be able to scale up the capacity to the levels required by predicted traffic demands~\cite{cisco2016}. Furthermore, moving data to remote cloud services might have serious privacy implications. This is the case, for example, of Industry 4.0 applications, where data generated in factory environments need to stay on devices owned by the manufacturing company for confidentiality reasons \cite{pppeu}. Similar concerns also arise in mobile social networking applications, whereby users might not be willing to send data over external remote cloud operators. 

Because of the above concerns, edge and fog computing solutions are gaining momentum~\cite{GarciaLopez:2015:ECV:2831347.2831354,Habak}. In edge and fog computing, data-centric services are provided, possibly in a cooperative way, by devices at the edge of the network, close to where the data are generated. This addresses the ``data-gravity argument'', according to which the huge amount of data generated at the edge of the network ``attracts'' computation towards the edge, and not vice-versa~\cite{sap2017}.

Along this line, in this paper we target \emph{service provisioning in mobile clouds of devices at the edge of the network}, using the opportunistic computing paradigm~\cite{Conti2013,Conti2010}. By mobile cloud, we denote a group of mobile devices at the edge of the network, moving in a limited physical area (whose dimension may range from an office environment to a region of a city). According to the opportunistic computing paradigm, mobile devices use each other's hardware and software resources when they are in physical proximity, to provide services to each other (e.g., to elaborate local data they have generated). The users' mobility enables direct device-to-device (D2D) opportunistic networking data transfers, which are used to exchange input/output service parameters, as well as to collect the information necessary to nodes to decide how to provide services to each other without centralised control.

This approach is relevant for several applications. As an example, in case of industrial applications, it allows, the owners of a factory to optimise the utilisation of the mobile devices used by the workers, and use remote cloud services only when needed. This can help reduce the cost of IT services, which is considered a significant benefit, in particular for small and medium enterprises (SMEs)~\cite{pppeu}. Moreover, this approach perfectly fits the scenarios for which Proximity-based Services (ProSe) are foreseen in long-term evolution (LTE) standards and 5G networks~\cite{3GPP2016}, whereby services relevant for a specific geographical area are directly provided by local nodes through D2D communications. Moreover, service provisioning in mobile clouds is also a viable option when users' devices store privacy-sensitive data that users would not share with third parties. For example, users' devices can provide data-centric services based on elaboration on their own data, without ``giving away'' raw data to third parties. Last but not the least, opportunistic computing supports service provisioning in case of infrastructure unavailability, such as incidents, disasters, censorship, or severe congestions.

In this paper, we assume that mobile nodes\footnote{The terms node and devices are used interchangeably in the paper.} request services that have to be provided by other nodes in the same mobile cloud. Nodes generating requests are hereafter referred to as \emph{seekers}. Given a particular service request, there can be multiple ways of providing it, such as entirely by a single device, or through a composition of service components, each provided by a different device\footnote{For the purpose of this paper, without loss of generality, \emph{sequential} components provided by the same device, can be merged together and considered as a unique service component.}. Hereafter, devices providing services (and service components) are referred to as \emph{providers}. Moreover, it if does not cause confusion, we will use the generic term ``service composition'' to indicate both single-component and multiple-components services, the former corresponding to a service composition of length one.

In particular, this paper proposes the \emph{Minimum Expected Value (MEV)} policy for service provisioning in mobile clouds. Upon a service request issued by a seeker, MEV aims at identifying the service provisioning option that minimises the expected time required by the seeker to obtain the results of the requested service (hereafter referred to as the \emph{service provisioning time}). To this end, nodes use a local view of the network, built through context information exchanged during opportunistic contacts with each other. Specifically, MEV estimates the expected service provisioning time of each composition known at the seeker (using local knowledge only) through an analytical model. The composition with the lowest expected service provisioning time (according to this model) is invoked. As explained in more detail in Section~\ref{sec:MEV-algorithm}, MEV could be easily adapted to the case of parallel executions for reliability. However, as the main focus of this paper is on service provisioning rather than on reliability (see Section~\ref{Related} for the rationale), we hereafter do not consider parallel service provisioning.

It is worth pointing out that, the main objective of the analytical model at the basis of MEV is not to predict the exact value of the expected service provisioning time of each possible composition satisfying a given request. Rather, the model aims primarily at \emph{ranking} service provisioning alternatives, in order to identify the one that will yield minimum service provisioning time. We anticipate that, even though not being its main objective, the model is also capable of predicting with quite good precision the actual average service provisioning time.

We assess the performance of MEV through simulation experiments. Specifically, we compare MEV with three alternatives (called AFIR, RAN, and ATO, defined in Section~\ref{sec:results}), which are representative of the state-of-the-art. We use TheOne~\cite{Keraenen2009}, which is a reference simulation environment  for opportunistic networks. Moreover, the simplifying assumptions used to derive the analytical model do \emph{not} hold in any of the simulations. To make our simulation results representative of a wide range of application scenarios and environments,  we use a number of real mobility traces, which exemplify a diverse set of human mobility patterns, from more ``compact'' scenarios where the nodes meet more frequently, to more sparse scenarios where nodes meet seldom, implying contact events are very rare. Moreover, we also vary (i) the size of the I/O parameters, i.e., the network congestion level; (ii) the contention at providers' CPU, i.e., the computation congestion at providers; and (iii) the service execution time, i.e., the computational complexity of service executions.

We validate the model accuracy in ranking service provisioning alternatives. Results show that the model is able to rank service compositions effectively (see Figure~\ref{fig:MEV-alternatives} in Section~\ref{sec:results}). Thanks to this ranking effectiveness, in between 60\% and 70\% of the cases, MEV selects a better service composition than the other three policies. Such a high precision makes MEV the policy which achieves the lowest average ``loss'', where loss is measured, for a given policy and a given service request, as the difference of service provisioning time between that policy, and an ideal (infeasible) policy that selects the policy (among all four) that \emph{will} yield the minimum service provisioning time.  After assessing the \emph{usefulness} of our analytical model in raking service compositions, we directly compare the performance of MEV against the other policies over the entire range of considered parameters. In all of these cases, our results show that MEV yields significantly lower service provisioning times with respect to the other three policies. Specifically, we find that the more scarce are the network or CPU resources, the higher is the advantage of MEV over the other policies. This shows that our model is particularly helpful for resource-limited devices, which is particularly appropriate for our reference scenarios. Moreover, we also show that MEV yields lower service provisioning times with respect to an ideal, but infeasible, policy, which, for any given request, adopts the policy, among AFIR, RAN, and ATO, that will yield the lowest service provisioning time.

We remark that, the fact that none of the simplifying assumptions used to derive the analytical model hold in the simulations, means that the gain of MEV over the other policies does not depend on these assumptions, and that the model at the basis of MEV is effective in ranking service provisioning alternatives irrespective of those assumptions. Moreover, the wide range of parameters considered gives generality to the superior performance of MEV with respect to the other policies.

The paper is organized as follows. Section \ref{Related} describes the main approaches presented in the literature for service provisioning and composition in mobile environments.  The MEV policy is described in \ref{sec:MEV-algorithm}, while Section \ref{sec:model} presents the analytical model that estimates the expected service provisioning time of a given service composition, on which MEV is based. In Section \ref{sec:results} we characterise the precision of the model in ranking service composition alternatives, and compare MEV with the three alternative policies. Finally, Section \ref{Conclusions} draws the main conclusions of the paper.

\vspace{-12pt}
\section{Related work}
\label{Related}

A first body of related
  work~\cite{Kalasapur2007,Bianchini2005,DelPrete2008,Wang2011} proposes
  service discovery and composition in mobile networks with stable connectivity
  where disconnections are a failure state or, equivalently, where service
  provisioning occurs only among devices in direct communication~\cite{Huerta-Canepa2010}.
  In this paper we focus on
  opportunistic networks where disconnections during service provisioning
  are the norm, and the entire service composition design is modified accordingly.

Service composition in dynamic mobile networks has been addressed also
  in~\cite{Tamhane2012,Sadiq2015}, by proposing simple heuristics for selecting
  compositions, while in this paper we exploit an analytical model to
estimate the lowest service provisioning time.
Parallel service execution in mobile clouds has been tackled in
  \cite{Tamhane2012,Passarella2011,Shi2012_SER,groba2014}. However, when
  composition is used~\cite{Tamhane2012,Shi2012_SER,groba2014} no model is derived to estimate
  the best composition, while when such a model is
  used~\cite{Passarella2011} service composition is not used. The main target of
  this paper is to improve over solutions in the first class, by deriving a model of service
  provisioning time in case of composition. As explained in
  Section~\ref{sec:MEV-algorithm}, the same approach and algorithm (with minimal
  modifications) can also be used in the case when multiple compositions
  obtaining the same composite service are run in parallel to improve
reliability of composed service execution.

Serendipity~\cite{Shi2012_SER} (also part of Cirrus
Clouds~\cite{Shi2012}) is an opportunistic service provisioning system that
shares some similarity with our proposal. However, in Serendipity the
composition of a service is given and fixed, and the focus is on (i)
prioritising among components executions, and (ii) optimising the completion time of each
individual component. Differently from our proposal, Serendipity can decide to
execute on providers not in contact with the seeker only in presence of a
coordination control channel thorough which providers' resources are
pre-allocated. In absence of such channel, which is the case we consider,
Serendipity greedily executes components on the first
encountered provider if it improves over local execution, while MEV
picks providers based on estimations of (i) their load, (ii) their service queue and (iii) the time to exchange I/O parameters. Note that using estimates for these parameters, as well as the time to encounter them, is the most important feature of MEV, which makes it significantly different from Serendipity. In Section~\ref{sec:results} we compare our proposal with AFIR, which works very similar to how Serendipity would work in our settings. The significant performance gap between the two (MEV achieves up to 43\% lower service provisioning time) shows that the different features included in MEV provide, in relevant scenarios, a significant benefit over Serendipity.

Other related papers touch upon orthogonal aspects with respect to the focus of
this paper. FemtoCloud~\cite{Habak} relies on a localised controller and a group of nearby
mobile devices available to share computing resources. The
controller centrally solves an
optmisation problem to allocate tasks to mobile devices while they are in range,
so as to
optimise the total number of useful computations in the group. In Mobile Device
Clouds~\cite{Mtibaa2013a,Mtibaa2013} mobile nodes allocate tasks to each other
to optimise the total energy consumption, and achieve the longest possible
time until the first node depletes energy. In case of opportunistic contacts, tasks can be
executed on providers only if contacts are long enough to complete their execution. Neither
FemtoCloud nor MDC consider service composition. Orthogonal to our work is
also~\cite{Mtibaa}, where an architecture is proposed to accomodate mobile cloud
service execution, leaving open the specification of the algorithms to be run
within the architecture. MEV could be easily plugged into this architecture.

MEV has some conceptual resemblance with utility-based forwarding in opportunistic network, such as, for example, Delegation Forwarding \cite{Erramilli2008}. However, with respect to Delegation Forwarding, MEV does not choose more than one service provider in parallel (which would be the equivalent of using multiple forwarders in Delegation Forwarding) and therefore the selected provider cannot be the one with the "max utility over those seen so far" as in Delegation Forwarding, but needs to be selected based on estimates of future contacts with suitable providers.

Finally, in this paper we extend our previous work  \cite{Conti2013,Mascitti2014}
where we provided the initial definition of MEV. With respect to \cite{Conti2013, Mascitti2014}, in this paper we
significantly refine the analytical model at the core of MEV to better take into
account the specificities of service composition. Moreover, we provide an extensive new set of performance results, showing that our analytical model is able to correctly rank service provisioning alternatives, thus making MEV more effective than the other considered policies. We also compare MEV with
these policies using real mobility traces, and discuss the advantages of the
proposed solution. Finally, we highlight the performance of MEV (in comparison with the other policies) by varying several parameters, including size of I/O parameters, network and providers' CPU contention, service execution time, mobility patterns.

\vspace{-12pt}
\section{The MEV algorithm}
\label{sec:MEV-algorithm}

\begin{figure}[t]
\centering
\subfloat[Service request resolution]{\includegraphics[width=.4\textwidth]{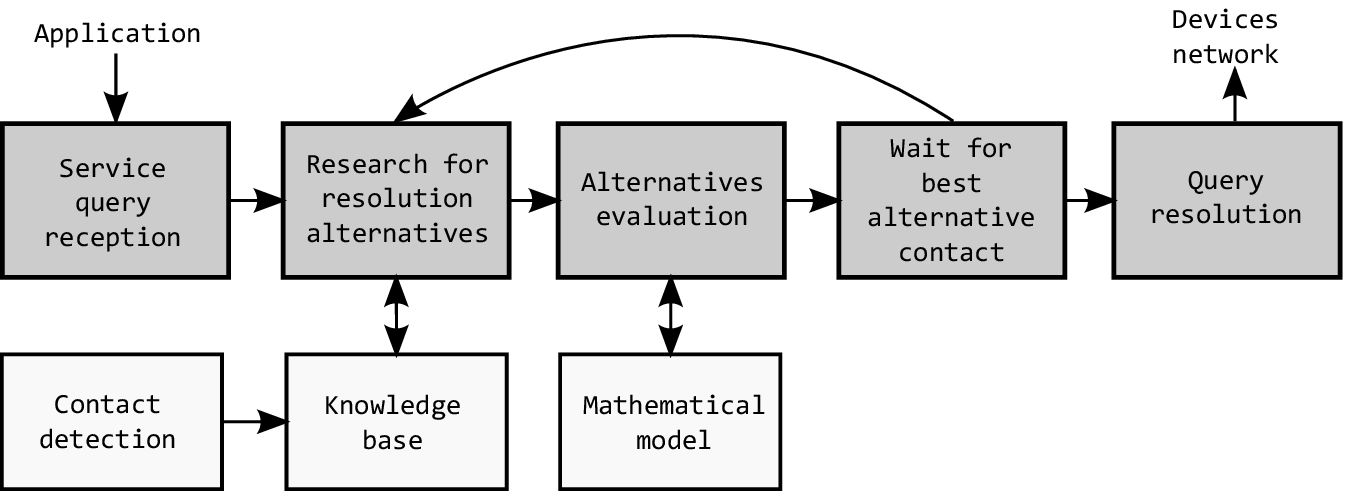}
\label{fig:risol}}
\quad
\subfloat[The Service Graph]{\includegraphics[scale=0.7]{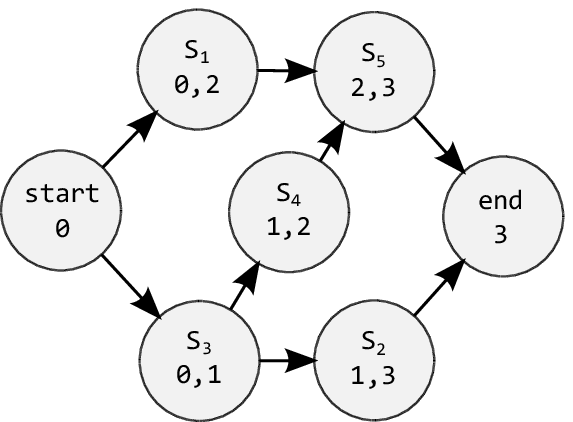}
\label{fig:ch5comp1}}
\subfloat[The Composition Graph]{\includegraphics[scale=0.45]{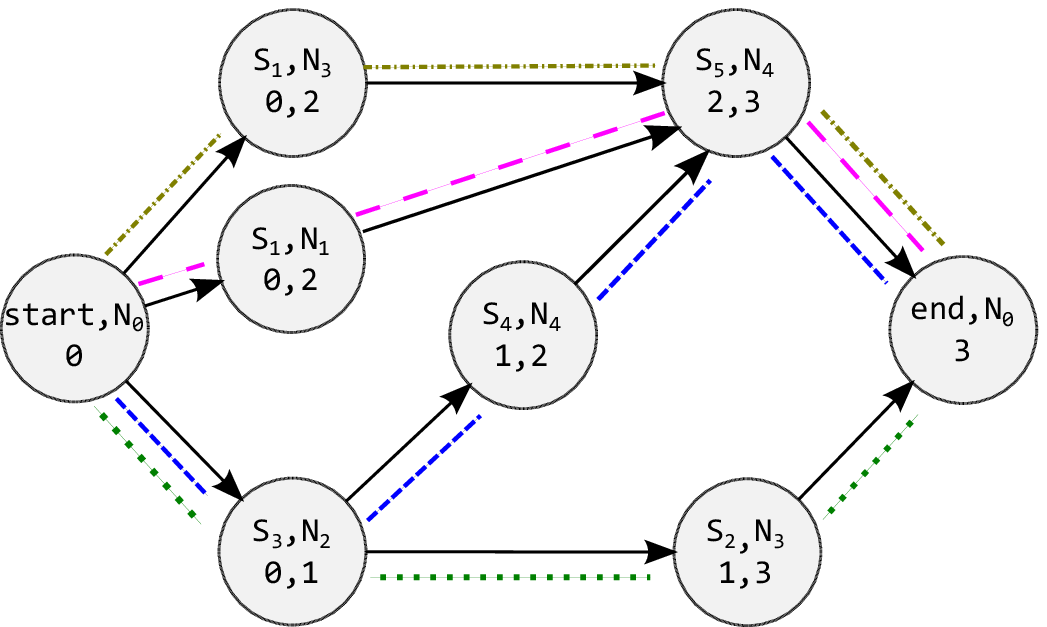}
\label{fig:ch5comp2}}
\caption{System algorithm and composition representation}
\vspace{-18pt}
\end{figure}

To describe the system behaviour we describe how a service request is managed and what is the logical decision process to select the components that form the composition involved in the resolution of the request. Fig. \ref{fig:risol} shows a graphical representation of the algorithm used by seekers.  Let us consider a tagged seeker running an application that at some point generates a request. The system running at the seeker first finds, through the knowledge base (i.e., a local storage with information about available services, managed as described in the following of the section), all possible compositions that would satisfy the request. Then, the service provisioning time of each composition is estimated, and the one providing the minimum provisioning time is selected. If the first provider in the composition is in contact with the seeker, the execution of the first component starts. Otherwise, the seeker waits to encounter this provider. In the meanwhile, it may encounter other nodes, which, as explained in the following, could result in updating the local knowledge at the node. In this case, the selected composition is re-evaluated, and possibly modified according to more refined knowledge acquired by the seeker. Eventually, the service composition starts and proceeds until the application request is satisfied. Between two consecutive components, output parameters of the former are passed to the next provider as input parameters.

Let us discuss more in detail how the key components of Fig. \ref{fig:risol} are realised. Let us first consider the block indicated as "research for resolution alternatives": once a service request is generated at a seeker, the system searches for resolution alternatives to satisfy the request. To this end, the system searches in the local knowledge base, in order to collect all the known service components that might be used and the statistics needed for the evaluation of the alternatives. The knowledge base is updated by each node upon encountering other nodes. In particular, for each encountered node it stores $(i)$ the set of provided services, $(ii)$ an estimate of the computation time for each provided service, $(iii)$ an estimate of the number of service requests generated by the node upon encounter, and the respective size of exchanged parameters, $(iv)$ an estimate of contact and inter-contact times, and $(v)$ an estimate of the average throughput available when the two nodes encounter. Specifically, for the reason explained in Section \ref{sec:model}, upon encountering, nodes exchange the first two moments of service computation time, as well as the list of provided services. The number and size of requests, contact and inter-contact times and average throughout estimates are simply monitored by each node without requiring exchange of information. This information is sufficient for the tagged node to estimate the service provisioning time, as we explain in Section \ref{sec:model}.

After collecting the service components from the knowledge base, the system builds a Service Graph out of them (an example is shown in Fig. \ref{fig:ch5comp1}) where vertices are components, and edges represent the fact that two components can be executed sequentially.  Note that in the Service Graph there is not yet information about which nodes provide components, which is added in the following logical step (the Composition Graph). Each path connecting two vertices of the graph shows a possible composition to satisfy the application request. Each service component $s_j$ is identified in our system as a pair $(I_j,O_j)$ where $I_j$ is the input type and $O_j$ is the output type of $s_j$. For the sake of simplicity, in the following we assume that these types are codified by integer values, furthermore we will consider acyclic compositions, i.e. compositions where the same components cannot appear twice. To ensure this, any Service Graph we consider will contain only services $s_j$ such that $I_j < O_j$. For instance, Fig. \ref{fig:ch5comp1} shows a set of service components $\{S_1, S_2, S_3, S_4,S_5\}$ linked by their type dependencies together with two special components $Start$ and $End$, representing the start and the end points of the service composition corresponding to the considered request. Note that, nodes $Start$ and $End$ do not correspond to any service component, and are only used to indicate the start and end of the service composition. This means that, in the composition graph, a single-component service (e.g., for service $s_i$ provide by node $n_j$) corresponds to three nodes, $(Start, n_0)\rightarrow(s_i,n_j)\rightarrow(End,n_0)$.

Given that each component may be offered by different providers known by the seeker, there can be different composition alternatives depending on the chosen providers. To identify these alternatives, a Composition Graph is created (as in Fig. \ref{fig:ch5comp2}), where, for each component, vertices are created for all known providers offering the component. On the resulting graph, each path from component $Start$ to component $End$ is a suitable composition. The graph is weighed, and weights are the key elements provided by our analytical model to estimate the service provisioning time, as explained in Section \ref{sec:model}. Note that the graph may change from node to node, as we assume it is built based on information available locally and collected through direct pairwise contacts, and not on global information. 

In the "alternatives evaluation" block, the list of alternatives taken from the graph is then evaluated by computing the estimated service provisioning time when using each possible composition. How the model does these estimates is described in Section \ref{sec:model}.
After the evaluation, the system ranks the alternatives, choosing the one with the least expected service provisioning time.

In the "wait for best alternative contact" block, the seeker, if it is not currently in contact with the chosen provider for the first service component, waits for a contact with it. Otherwise the decision process is over and the service request is queued to be sent to the provider. If the seeker has to wait for the first provider, it continues to monitor the state of the network upon new contacts. Any new contact triggers a new exchange of information between the nodes. This information may alter the classification computed previously, so, in this case, the system goes back to the alternatives evaluation phase to update the ranking of the compositions.

\vspace{-12pt}
\section{Modelling service provisioning time} 
\label{sec:model}
This section introduces the stochastic model exploited to estimate the service provisioning time. Without loss of generality, we present the model by focusing on the provisioning time of a service requested by a tagged seeker. For the sake of explanation, let us first assume that this service can be provided directly by a tagged provider encountered by the seeker (i.e., the service composition is made of 1 service component only). We then extend it to the general case of composed services. Note that, in the following, when required, we denote with $h$ the index of the tagged seeker, and $j$ the index of the tagged provider. Most of the variables used in the model refer to the pair (tagged seeker, tagged provider). When this dependence is clear, we omit to use the indices $h,j$, to make the notation simpler. The service provisioning time in this case (hereafter denoted with $R$) is made up of five consecutive phases, as follows: 
\begin{itemize}
\item{\em Contact of the service provider} ($W$). This is the time needed by the seeker to encounter the provider after the point in time when the service request is generated. It is determined by the inter-contact time between the seeker and the provider.

\item{\em Data transfer} (Input Time $B$, Output Time $\theta$). This is the time needed to transfer the input parameters from the seeker to the provider and the output parameters from the provider to the seeker (after the execution time is complete). In the latter case, we also include in this phase the time required by the provider to meet the seeker from the point in time when the service execution is complete. Note that, in an opportunistic network data transfers between two nodes  may be affected by connection disruptions, due to the nodes mobility, but also by transfer from and to other nodes that use the same shared medium, or even by other concurrent transfers between the seeker and the provider involving other requests.
\item{\em Queue waiting time} ($DQ$). Once onto the provider, actual execution may be delayed due to previous pending requests. We model this as a FIFO queue at the provider. The duration of this delay depends both on the frequency of the request arrivals on the provider and on the time to process them.
\item{\em Service execution time} ($DS$). This is the time to execute a service on the provider. It depends on both its computational capabilities and on the type of the service. 
\end{itemize}
Each of these phases can be modeled as a separate random variable as we will describe in the following of the section. Since they are sequential, we obtain for a single component service the following expression:
\begin{equation}
\label{eqn:esingle}
 R_{single} = W+B+DQ+DS+\theta 
\end{equation}
For the case of a service composition made of $n$ components, the service provisioning time can be expressed as follows:
\begin{equation}
\label{eqn:ecomp}
R_{comp} = W+B+\sum_{i=1}^{n}( DQ_i+DS_i+\theta_i)
\end{equation}
In Equation \ref{eqn:ecomp}, $W$ and $B$ refer to the time to meet the first provider in the composition and transfer the input parameters to it, respectively. $DQ_i$ and $DS_i$ are the length of queuing at the $i$-th provider and executing the $i$-th component, respectively. $\theta_i$ represents the time required by provider $i$ to encounter provider $i+1$ (or to encounter the seeker, for provider $n$), and transfer to it the output parameters of component $i$ which become input parameters of component $i+1$ (or, for provider $n$, to transfer the final output parameters to the seeker). 

In the following of the section we explain how to estimate the expected time taken by these phases. We discuss the derivation of $B$ and $\theta$ last, as this requires a number of steps. Through these estimates and using Equations \ref{eqn:esingle} and \ref{eqn:ecomp} each seeker can estimate the expected service provisioning time of the available compositions, and pick the best one.

\subsection{Contacting the service provider}
Random variables $T_{C}$ and $T_{IC}$  model, respectively, the contact and inter-contact times between two nodes $h$ and $j$.  For each pair of nodes, we assume that contact and inter-contact times between those nodes are independent and identically distributed (i.i.d.). We also assume that contact and inter-contact times of different pairs of nodes are independent of each other. Finally, we assume that the variables $T_{C}$ and $T_{IC}$ follow exponential distributions with rates $\delta$ and $\delta'$. As shown by real trace analysis presented, for example, in \cite{Gao2009,Tournoux2011}, although controversial, exponential contact and inter-contact times is one of the possibilities, and is a common assumption in the literature on opportunistic networking and computing (e.g. \cite{Passarella2011,Spyropoulos2009}). Also note that analysis on real traces in Section~\ref{sec:results} (where non-exponential contact and inter-contact are also present) show that the model works well also in more general settings. Since a node cannot know beforehand the values of $\delta$ and $\delta'$, each node computes an estimate of these values by averaging the values of contact and inter-contact times with other nodes collected during opportunistic contacts.

The time for seeker $h$ to contact the generic service provider $j$, is denoted by the random variable $W$.  This is equal to 0 if, at the time when the evaluation is done, $h$ and $j$ are in contact, while it is equal to the residual inter-contact time $T_{IC}$ otherwise (and, under our assumption, its expected value is equal to $E[T_{IC}]$ due to the memoryless property of the exponential distribution).
The expected value of $W$ is therefore:
\begin{equation}
E[W] =
  \begin{cases}
   \frac{1}{\delta'} & \text{if $h$ and $j$ are not connected} \\
   0       & \text{otherwise }
  \end{cases}
\end{equation}
\subsection{Service execution time}
The random variable $DS_{s_i,j}$ for the time needed to execute service component $s_i$ on a provider $j$ is influenced both by the device computational capabilities and by the implementation of the requested service component. We assume that each provider keeps an estimate of the expected value of $DS_{s_i,j}$ based on the previous local executions, and sends this value to the encountered nodes. Similarly, providers also keep an estimate of the second moment of $DS_{s_i,j}$ and send it to other nodes during contacts. This is used in the estimation of the waiting time $DQ$, as explained in the next section. It is worth discussing why we chose to use the \emph{measured} first two moments of the execution time, instead of defining an analytical model for this component. In general, service execution times will depend on a large number of parameters, also including the impact of other local applications running on the device. Correctly modelling these factors would have made our model too complex to be practically useful. Also note that, the goal of our model is not to describe precisely this part of the service provisioning process, but showing that, if we can measure simple statistics about it (the first two moments), we can build an efficient algorithm to enable opportunistic service computing.

\subsection{Queue waiting time}
We assume that the seeker locally generates requests addressed to the provider according to a Poisson process with rate $\lambda$. Therefore, when the provider is encountered, in general a batch of requests may have been accumulated, whose input parameters need to be transferred. To account for this, we model the service provider with a $M^{[X]}/G/1$ queue. As shown in \cite{Takagi1991}, this is exact when $(i)$ requests are generated according to a Poisson process, $(ii)$ inter-contact times are exponential, and $(iii)$ all requests stored at the seeker are transferred to the provider during a contact. Under our assumptions, hypothesis (iii) may not hold. However, we still use the $M^{[X]}/G/1$ model, and assess the approximation level by validating the results obtained using this model against simulations in Section \ref{sec:results}.

Let us denote with $L$ the number of  requests in a batch received by the provider.
The expected value of the random variable $DQ_{j}$  can be computed  \cite{Takagi1991} based on the first two moments of the service execution time $DS_{s_i,j}$ (with expected value $d$, second moment $d^{(2)}$ and average service rate $\mu$) and of the random variable $L$ (with expected value $l$ and second moment $l^{(2)}$). These values can be estimated by monitoring the batches arriving to the provider and the executed services. In addition, the provider estimates the rate $\lambda$ of the request batches and computes the average load $\rho$ of the provider as $\lambda*l*d$.
Starting from these values, the expected value of $DQ_{j}$ can be computed, as shown in \cite{Takagi1991}, as:
\begin{equation}
E[DQ_{j}]=\frac{\lambda l d^{(2)}}{2(1-\rho)}+\frac{l^{(2)} d}{2l(1-\rho)}
\end{equation}
\subsection{Data transfer times}\label{sub:out}
\begin{figure}[t]
	\centering
	\includegraphics[scale=0.75]{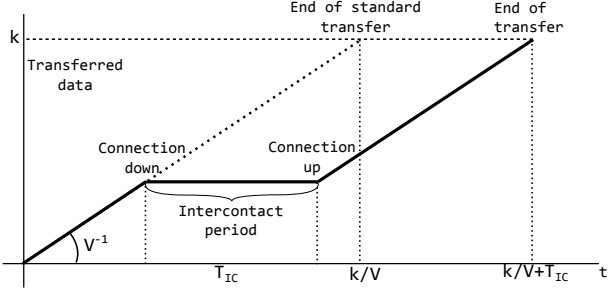}
	\caption{Modelling Data Transfer}
	\label{fig:upload}
\end{figure}

Unlike most analytical models in opportunistic networks, we assume that the throughput available to nodes during contact times is finite. Therefore, in our model we need to take into account the possibility that data transfer may be interrupted by disconnections, and therefore needs to be resumed at the next contact event. In the following, we denote with $V$ the average throughput experienced between the tagged seeker and provider and assume it is estimated by the two nodes through a conventional smoothed average algorithm using the throughput samples obtained during actual data transfers between them. We also denote with $k$ and $k'$ the size of the input and output data to be transferred for the requested service. Fig. \ref{fig:upload} shows an illustrative example where one disconnection occurs during the transfer of the input data of size $k$. In addition to the time needed to actually transfer data between the nodes, we have to take into account additional inter-contact times between consecutive contact events. 

First of all, we analyse the case of data transfers in single-component services. In this scenario we identified three cases to model depending on the time when the first contact (that we refer as $T_C(0)$)  used to transfer data ends. The first case (\emph{case 1}) is a scenario where all the phases of the execution can be completed without any interruption. In the second case (\emph{case 2}), only the first data transfer (the one used to transfer the input data of size $k$ for the service execution) completes without interruptions, while there is at least one disconnection before the seeker completely receives the result of the service execution. In the last case  (\emph{case 3}), there is at least one disconnection before the input data transfer phase is completed.

In the following we provide the expressions for $B$ and $\theta$, considering their formulations involving seeker $h$ and provider $j$. Specifically, we analyse cases 1 and 2 in the paper. Case 3 is conceptually analogous to case 2, and therefore we include it in the Appendix\footnote{The Appendix is provided as supplemental material of this submission.}. Then, we model the probabilities of the cases, i.e. $p_1$, $p_2$ and $p_3=1-p_1-p_2$. Note that $B$ is the same in cases 1 and 2.  The expected values of $B$ and $\theta$ can be immediately derived by applying the law of total probability:
\begin{equation}
\label{btotsingle}
E[B]=E[B | case 1,2]*(p_1+p_2)+E[B | case 3]*p_3
\end{equation}
\begin{equation}
\label{ttotsingle}
E[\theta]=E[\theta | case1]*p_1+E[\theta | case2]*p_2+E[\theta | case3]*p_3
\end{equation}
\subsubsection{Analysis of case 1}
In  \emph{case 1}, with no interruptions, the input and output transfer times for a request from seeker $h$ for a service provided by node $j$, depend on the throughput available between the nodes $V$, the sizes of the input and output parameters ($k_{data}, k'_{data}$ respectively, whose value depends on the requested service) and the size of the queued data that has to be transferred between $h$ and $j$ before the input and the output data transfers can be started (respectively called $k_{queue}$ and $k'_{queue}$). The value of $k_{queue}$ can be directly observed by the seeker during the evaluation of the alternatives, while $k'_{queue}$ is estimated as the average size of queued data in $j$ addressed to $h$ at the end of service executions.
Therefore, $B$ and $\theta$ for \emph{case 1} can be modeled as follows: 
\begin{equation}
\label{eqn:ecase1}
B|case1=\frac{k}{V} \quad \theta|case1=\frac{k'}{V}
\end{equation}
Where $k=k_{data}+k_{queue}$ and $k'=k'_{data}+k'_{queue}$.
These values of $\frac{k}{V}$ and $\frac{k'}{V}$ can be called the \emph{net transfer time} for $B$ and $\theta$, as they are the minimum estimated data transfer times without the presence of any interruption.

\subsubsection{Analysis of case 2}
In the second case (\emph{case 2}) the first data transfer can be completed during the first contact, but the rest of the process cannot. The input time $B$ is the same already analysed in Equation \ref{eqn:ecase1}. 

We analyse $\theta$ by considering the two possible cases: the case where the (first) contact\footnote{Here first contact still refers to the contact between the seeker and provider at the beginning of the input data transfer phase. It is not the first contact during which the output data transfer starts.} ends while the transfer of output parameters is already ongoing (case 2A), and the case when the contact ends before the output data transfer has started. In the first sub-case, the first contact can be used to start the output data transfer, but the same contact is not long enough to complete the entire transfer (otherwise it would fall in case 1). Otherwise, if the first contact ends before the service execution is completed, $\theta$ may start during a contact period (case 2B) or an inter-contact period (case 2C).

\begin{figure}[t]
\centering
\includegraphics[width=.48\textwidth]{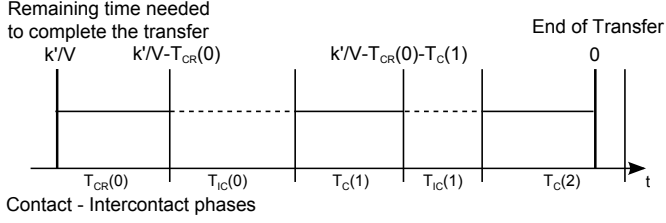}
\caption{Phases of an output transfer starting during a contact, with 2 disconnections afterwards} 
\label{fig:transfer}
\vspace{-12pt}
\end{figure}

$\theta$ in case 2A can be expressed as in Equation \ref{eqn:theta2A}. The key characteristics of case 2A are that $(i)$ $\theta$ starts during the first contact between the seeker and provider $(ii)$ then $\theta$ always includes the following inter-contact time ($T_{IC}(0)$ in Equation \ref{eqn:theta2A}), and $(iii)$ it then finally may include an additional number $N_{2A} \geq 0$ of inter-contact times. An example of case 2A can be seen in Fig. \ref{fig:transfer}, which represents an output transfer phase with 2 disconnections. As shown in the following, $N_{2A}$ can be characterised based on the number of contacts needed to transfer $k'$ bytes minus the data already transferred during the first contact. Considering that $\theta$ is made up of the net transfer time ($k'/V$) plus the additional inter-contact times needed to complete the transfer, it can be expressed as follows:
\begin{equation}
\label{eqn:theta2A}
\theta|case 2A = \frac{k'}{V}+T_{IC}(0)+\sum_{i=1}^{N_{2A}}T_{IC}(i)
\end{equation}
The characteristics of case 2B are that $(i)$ $\theta$ starts during a contact time , and $(ii)$ it may include a number $N_{2B} \geq 0$ of inter-contact times. As shown below, as we assume that contact times are exponential and thus memoryless, $N_{2B}$ can be characterised based on the number of contacts needed to transfer $k'$ data (this is why $N_{2B}$ is stochastically different from $N_{2A}$, as $N_{2A}$ is the time to transfer k' data minus what has been transferred already during the first contact). Therefore, $\theta$ in case 2B can be expressed as follows:
\begin{equation}
\label{eqn:eqnt2b}
\theta|case 2B = \frac{k'}{V}+\sum_{i=1}^{N_{2B}}T_{IC}(i)
\end{equation}
Finally, the characteristics of case 2C are that $(i)$ $\theta$ starts during an inter contact time, and $(ii)$ it may then include an additional number $N_{2C} \geq 0$ of additional inter-contact times. It is easy to see that, as we have assumed that contact times are memoryless,  $N_{2C}$ is stochastically equivalent to  $N_{2B}$. Therefore, by denoting with $T_{ICR}(0)$ the residual duration of the inter-contact time during which it starts, $\theta$ in case 2C can be expressed as follows:
\begin{equation}
\label{eqn:eqnt2c}
\theta|case 2C= \frac{k'}{V}+T_{ICR}(0)+\sum_{i=1}^{N_{2B}}T_{IC}(i)
\end{equation}
To find the expected values for $\theta$ in this three sub-cases, we need to derive the distributions of $N_{2A}$ and $N_{2B}$. Hereafter we provide an intuitive derivation of the former. All details can be found in the Appendix.

\begin{lemma}
\label{lemN2A}
The probability that, in case 2A, $\theta$ includes $N_{2A}=n$ additional inter-contact times (after the first one that it always includes) is as follows:
\begin{equation}
\label{eqn:eqN2A}
P\{N_{2A}=n\}=e^{-\delta\frac{k+k'}{V}}*\frac{(\frac{\delta k'}{V})^{n+1}}{n+1!}*\frac{(1-\rho)\mu}{\delta+\mu(1-\rho)}
\end{equation}
\end{lemma}
\begin{IEEEproof}
We denote by $T_{CR}(0)$ the part of the first contact time between the seeker and the provider, that is the initial part of $\theta$. Therefore, the remaining transfer time is $k'/V-T_{CR}(0)$. In order for $N_{2A}$ to be equal to $n$, the remaining transfer time must be longer than $n$ contact times but shorter than $n+1$ contact times. Equation \ref{eqn:eqN2A} follows, as shown in detail in Appendix \ref{appendix}. 
\end{IEEEproof}

The closed-form expressions of $E[\theta]$ in the three cases can be derived as follows (see the Appendix for the proofs).

\begin{lemma}
\label{lemcaseA}
The expected value of $\theta$ in case2A, 2B and 2C are equal to: 
\begin{multline}
E[\theta| case2A]=\frac{k'}{V}+\frac{1}{\delta'}+\frac{1}{\delta'}*e^{-\delta\frac{k+k'}{V}}*\\
*(e^{\delta\frac{k'}{V}}*(\delta\frac{k'}{V}-1)+1)*\frac{(1-\rho)\mu}{\delta+\mu(1-\rho)}\\
\end{multline}
\begin{equation}
\label{eqn:ecaseB}
E[\theta| case2B]=\frac{k'}{V}+E\left[\sum_{i=1}^{N_{2B}}T_{IC}(i)\right]=\frac{k'}{V}\left(1+\frac{\delta}{\delta'}\right)
\end{equation}
\begin{multline}
E[\theta| case2C]= \frac{k'}{V}+E[T_{IC}(0)]+E\left[\sum_{i=1}^{N_{2C}}T_{IC}(i)\right]=\\
=\frac{1}{\delta'}+\frac{k'}{V}\left(1+\frac{\delta}{\delta'}\right)
\end{multline}
\end{lemma}

To conclude the analysis of case 2 we need to derive the probabilities of the three subcases, 2A 2B and 2C, conditioned to the fact that we are in case 2\footnote{We derive the probabilities of cases 1, 2 and 3 later, in Section \ref{cases}}. The probability of case 2A is provided in the following Lemma:

\begin{lemma}
The probability of the first contact lasting enough to complete the input data transfer, but not enough to reach the beginning of the output data transfer, is:
\begin{equation}
\label{eqn:eqn20}
p(Case2A)=\frac{e^{-\delta\frac{k}{V}}*\mu(1-\rho)}{\delta+\mu(1-\rho)}
\end{equation}
\end{lemma}
\begin{IEEEproof}
This probability can be written as the probability that the first contact time $T_{C}(0)$ is longer than the time for the transfer of the input parameters ($k/V$) plus the queuing time at the provider ($DQ_j$) plus the service computation time ($DS_{i,j}$), but not long enough to also include the transfer of the output parameters ($\theta$). Equation \ref{eqn:eqn20} can be derived by exploiting this formulation, as shown in the Appendix. 
\end{IEEEproof}

The probabilities of sub-cases B and C are calculated using the complement of the probability of case A "weighted" with the steady state probabilities of contact and inter-contact phases. Specifically, case 2B corresponds to the case where the end of the service computation time falls in a contact time. For simplicity, we assume that this occurs with a probability equal to the steady state probability that a random point in time falls inside a contact. Analogously, case 2C occurs with the steady state probability that a random point in time falls inside an inter-contact. Therefore we obtain:
\begin{equation}
p_{case2B}=(1-p_{case2A})* \frac{E[T_{C}]}{E[T_{C}]+E[T_{IC}]}
\end{equation}
\begin{equation}
p_{case2C}=(1-p_{case2A})*\frac{E[T_{IC}]}{E[T_{C}]+E[T_{IC}]}
\end{equation}
Using this probabilities we can find the expected value for $\theta | case2$ through the law of total probability.

\vspace{-12pt}
\subsubsection{Probabilities of single-component-service cases}\label{cases}
Having derived all the components of the service provisioning time in all cases, we now need to derive expressions for the probabilities of the three cases in which we have divided the analysis. 

Remember that case 1 is the case when the entire service provisioning time R is shorter than the first contact time between seeker and provider. Therefore the probability of case 1 is $P\{R<T_{C}(0)\}$. The following lemma provides a closed form for this probability, and for $p_3$. Probability $p_2$ is derived as $1-p_1-p_3$. 

\begin{lemma}
\label{lemp12}
The probabilities $p_1=P\{R<T_{C}(0)\}$ and $p_3$ can be approximated as:
\begin{equation}
p_1=\frac{\mu (1-\rho)e^{-\delta \frac{k+k'}{V}}}{\delta + \mu (1-\rho)}
\end{equation}
\begin{equation}
p_3=P\{T_{CR}(0)<\frac{k}{V}\}=F_{T_{C}(0)}\left(\frac{k}{V}\right)=1-e^{-\frac{\delta k}{V}}
\end{equation}
\end{lemma}
\begin{IEEEproof}
See Appendix \ref{appendix} for $p_1$. For $p_3$, its expression is straightforward, thanks to the assumption that contact times are exponentially distributed, as it is is the probability that the residual of the first contact $T_{CR}(0)$ is shorter than the time $k/V$ needed to transfer the input data without interruptions.
\end{IEEEproof}

\subsubsection{Data transfer for service compositions}\label{composition}
To complete the analysis, we now extend the derivation of the service provisioning time to the case of multi-component services.   As discussed before, this is expressed by Equation \ref{eqn:ecomp}, i.e.:
$$R_{comp} = W+B+\sum_{i=1}^{n}( DQ_i+DS_i+\theta_i)$$
For $W$, $B$, $DS_i$ and $DQ_i$ we can use the same formulation used in the single service executions, as they have no differences. For each $i$, $\theta_i$ clearly depends on the seeker-provider pair corresponding to component $i$. Also in this case, to simplify the notation, without loss of generality, we omit the indices of the specific pair, and provide the expression of $\theta_i$ for a generic component $i$, that we simply refer as $\theta$. In the formulation of $\theta$, the first factor includes the time to encounter the next provider, and possibly the time needed to transfer previously queued data (e.g., other input/output parameters) to be exchanged between these two nodes. For simplicity, we keep this factor as a model parameter (called \emph{average transfer queue time} $TQ$), and we assume that nodes estimate its value by monitoring previous data transfers between the same nodes through a standard smoothed average estimator. This is a well-established and solid estimator, which is able to capture also fluctuations of the estimated figure. On the other hand, the second part is analogous to $\theta$ for single service composition in case 2B, i.e. when it starts at the beginning of contact\footnote{More precisely, in case 2B, $\theta$ starts \emph{during} a contact. However, the average values become the same due to the assumption of contact times being exponentially distributed}, and therefore can be expressed as in Equation \ref{eqn:eqnt2b}. Therefore, $\theta$ becomes as follows:
\begin{equation}
\theta=TQ+\frac{k'}{V}+\left[\sum_{i=1}^{N_{2B}}T_{IC}(i)\right]
\end{equation}
Given that $TQ$ is a non-random parameter, we can calculate the expected value of $\theta$, using again the results provided in Lemma \ref{lemcaseA}, as:
\begin{equation}
E[\theta]= TQ +\frac{k'}{V}\left(1+\frac{\delta}{\delta'}\right)
\end{equation}

\subsection{Weights of the Composition Graph}\label{Composition}

The model presented so far allows us to compute all possible service provisioning times for all alternatives. To this end, we use the Composition Graph (Fig. \ref{fig:ch5comp2}) and (i) weigh the edges of the graph using the appropriate parts of the analytical model described in Section~\ref{sec:model} (the details on how this is done are presented in the following of the section), (ii) use a standard shortest path algorithm to find the alternative with the minimum estimated service provisioning time.

Remember that any edge $(s_i,n_j)(s_k,n_h)$ in the Composition Graph means that it is possible to compose services $s_i$ (provided by node $n_j$), and $s_k$ (provided by node $n_h$). For any such edge, its weight $\omega$ is the expected time between the end of the execution of service $s_i$ on node $n_j$ and the end of the execution of service $s_k$ on node $n_h$, both for single service execution and sequential compositions.  We can identify three types of edges that need different types of weights:
\begin{itemize}
\item{Starting edges}: The edges outgoing from the $start$ node, represented as $(start,n_j)(s_k,n_h)$, must be weighted with the estimated time to wait for the next contact with the provider, plus the time to transfer the input data for the service provisioning, the queue waiting time and the service component execution time, obtaining $\omega((start,n_j)(s_k,n_h))=E[W_{n_j,n_h}+B_{n_j,n_h}+DQ_{n_h}+DS_{s_k,n_h}]$
\item {Ending edges}: The edges incoming to the $end$ node, represented as $(s_i,n_j)(end,n_h)$, are only weighted with the estimated time  $E[\theta_{n_j,n_h}]$ to transfer the output of the service provisioning to the seeker. We have to consider that the formula is different for single-component and multiple-component services. Remeber (Section~\ref{sec:MEV-algorithm}) that in the composition graph a single-component service corresponds to three nodes, i.e., $(Start, n_0)\rightarrow(s_i,n_j)\rightarrow(End,n_0)$. In this case, the type $I_{s_i}$ (i.e., the input required by service component $s_i$) matches the output $O_{start}$ of node $start$. In terms of notation, we refer to $\theta$ in this case as $\theta_{n_j,n_h}$, and as $\theta C_{n_j,n_h}$ in the other case. $\theta_{n_j,n_h}$ and $\theta C_{n_j,n_h}$ are the weights on the ending edges in the two cases.
\item{Intermediate edges}: These edges $(s_i,n_j)(s_k,n_h)$ are only between two providers, so they are part of a composition. Their weight, similarly to the starting edges, is the sum of the estimated time $E[\theta C_{n_j,n_h}]$ to transfer data between the providers, the estimated queue waiting time $E[DQ_{n_h}]$ on the second provider and the estimated service execution time $E[DS_{s_k,n_h}]$ for service $s_k$.
\end{itemize}

\vspace{-12pt}
\section{Performance Evaluation}
\label{sec:results}

In this section, we first assess the precision of the model in ranking possible alternatives for service provisioning in Section~\ref{sub:ranking-precision}. Remember that providing an effective ranking of the known alternatives is the key objective of our model. Then, in Section~\ref{sub:comparison-policies} we compare the performance of the policy that uses our model to select the service composition to be used, with other reference alternatives, over a range of different parameters. While the first session mainly \emph{demonstrates the usefulness} of the model in identifying the best service composition, the second session mainly \emph{quantifies the effectiveness} of a policy using the model over a range of key parameters.

We define MEV (\emph{Minimum Expected Value}) as the policy that selects the service composition corresponding to the minimum expected service provisioning time according to our model. We consider three alternative policies: a Random policy (called \emph{RAN}) that makes a random selection among all service compositions known at the seeker; the Atomic policy (\emph{ATO}) that selects randomly a single-component solution (i.e., not using multiple-components service provisioning options); and the Always First (\emph{AFIR}) policy. AFIR always picks the first encountered provider that allows the composition to progress towards the complete composed service. Note that, in our case, as we consider sequential compositions only, AFIR is equivalent to Serendipity~\cite{Shi2012_SER}. Comparison with AFIR gives thus a comparison with one of the reference policies in the state of the art.

We run trace-based simulations using the PMTR~\cite{pmtr-2008}, Haggle~\cite{haggle-2009}, MIT-Reality \cite{mit-reality-2005} and Rollernet \cite{rollernet-2009} traces, which are reference traces in the literature. Among the available Haggle traces, we used the one related to the Infocom 2005 and 2006 experiments (throughout referred to as Info05 and Info06). Unless otherwise stated, simulations are repeated 5 times using independent seeds, and average results are presented with 95\% confidence intervals \footnote{Replicating more times did not yield to significantly lower confidence intervals}. We defined a warm-up period during which no requests are generated, and nodes only acquire information about contact and inter-contact times to yield stable estimates. Each service is identified by the type of its input/output, which is represented by an integer. In our simulations, input types {\it i} are selected in the range [0,7], while output types {\it o} in the range [1,8], with the constraint that {\it i} must be less than {\it o} to avoid cyclic compositions. The average service execution time for a service component is 75s, and service execution times are drawn from an exponential distribution. Using a random variable for service execution time, we take into account inevitable variability in service execution also for the same type of services (i.e., for a given value of their average execution time). Unless otherwise stated, service inter-generation time at seekers is uniformly distributed in the interval [40-80]s. We consider two sizes of input/output parameters, i.e., 40 KB and 1280 KB, representing respectively light and heavy data transfer cases. Finally, we take into account contention on providers' CPUs as follows. We consider that the CPU of providers is shared round robin among a maximum number of competing processes, $M$. The evolution of the competing processes over time is a birth-death process. Specifically, every quantum of time (0.1s in our simulations) the number of competing processes is increased or decreased by 1 with probability 0.1\%, simulating the activation and deactivation of processes. Parameter $M$ allows us to vary the average number of competing processes, and thus the level of contention on providers' CPUs. As far as network congestion is concerned, the maximum bandwidth provided by the underlying communication technology is split equally among neighbouring nodes with active communications.

whenever a node communicates during a contact with another node, we equally split the maximum bandwidth between all nodes in its communication range.

Note that in the following we present results obtained by varying most of these parameters over significant ranges. Therefore, in addition to comparing our policy with others, results in this section also show the effectiveness of our model to select the most performing service provisioning alternative also in scenarios where the modelling assumptions do not hold true. Specifically, in all simulations contact and inter-contact times are driven by mobility traces, and are not exponential, and service request generations are not Poisson. These are the two fundamental assumptions of our model: remember that for the other parameters (such as, e.g., the average service execution time, the average bandwidth between nodes) our model uses their average values monitored over time, and therefore there are no corresponding modelling assumptions.

\subsection{Ranking precision of the model}
\label{sub:ranking-precision}

\subsubsection{Precision of alternative policies}
\label{subsub:precision-policies}

We compare the four policies using two performance indices. We evaluate the percentage of service requests for which MEV is the best policy among the four, i.e., the policy selecting the composition with minimum service provisioning time. This index shows the effectiveness of MEV in identifying a better service composition with respect to AFIR, RAN and ATO. The second index is, for each policy, the average difference in service provisioning time between it and the best among the four. This index shows the average ``loss'' incurred by using each of the policies with respect to an ideal policy that for each service request takes the best among the options identified by the four policies.

For each request generated at a seeker, we run all four policies in parallel, to make sure that they all evaluate service compositions based on the same conditions. For fairness, in the simulations, the order in which the policies are scheduled is randomised, to make sure no policy obtains an unfair advantage due just to be scheduled first during contacts. A new service request is generated when the best composition among the selected ones has terminated. To have a number of samples of service provisioning time sufficient to compare the four policies, we replicated experiments 40 times (replaying the RollerNet trace several times to have a sufficiently long trace). To compute the loss index, we grouped requests in groups of 200 consecutive requests. We averaged the loss of each group of 200 requests, obtaining a sample of the average loss from each group. Finally, we computed the average value of the index over all groups, with 95\% confidence level. The simulation parameters are shown in Table~\ref{tab:ranking-sim-params}.

\begin{table}[ht]
	\vspace{-12pt}
	\caption{Simulation parameters for ranking experiments}
	\vspace{-12pt}
	\centering{
	\begin{tabular}{|c|c|}
		\hline 
		Trace & RollerNet \\ 
		Simulation Time & 100000s \\ 
		Warm-up time & 30000s \\ 
		Component avg. execution time & 75s \\ 
		Input/Outpup size & 40KB, 1280KB \\ 
		Avg. CPU contention (processes) & 3 (low), 10 (high) \\ 
		\hline 
	\end{tabular} 
	}
\label{tab:ranking-sim-params}
\end{table}

\begin{figure}[htb]
	\centering
	\includegraphics[width=0.24\textwidth,height=0.2\textwidth]{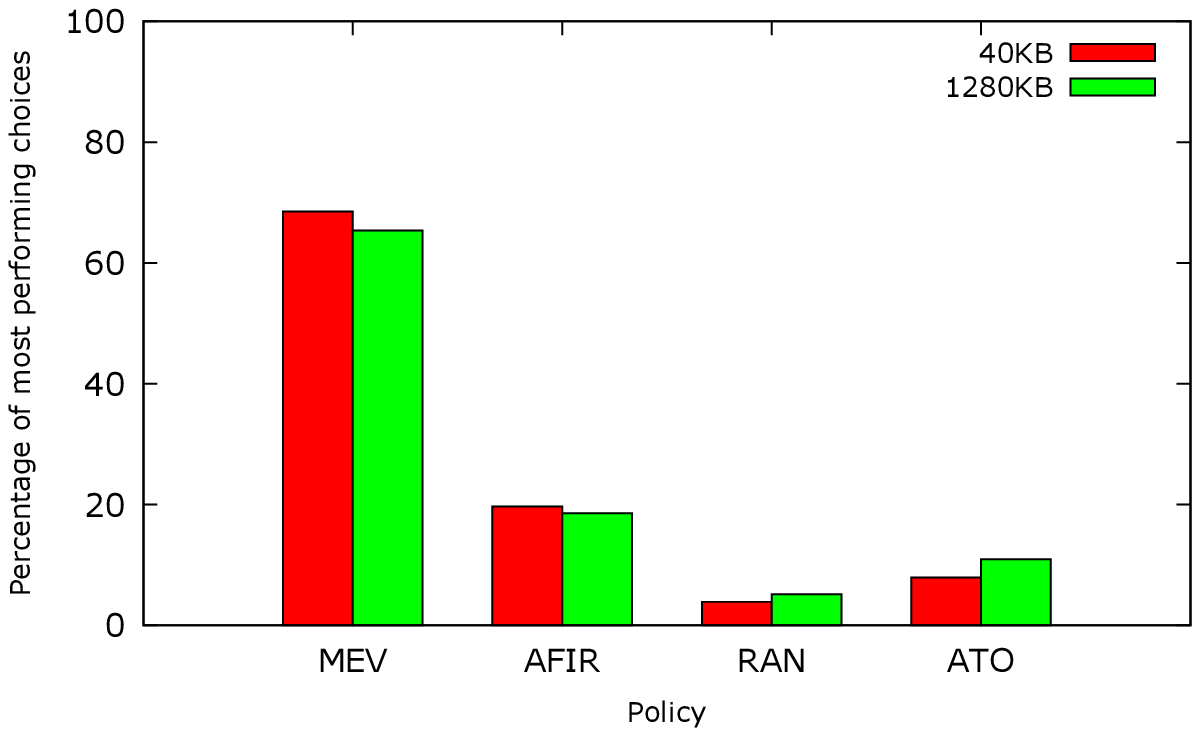}
	\includegraphics[width=0.24\textwidth,height=0.2\textwidth]{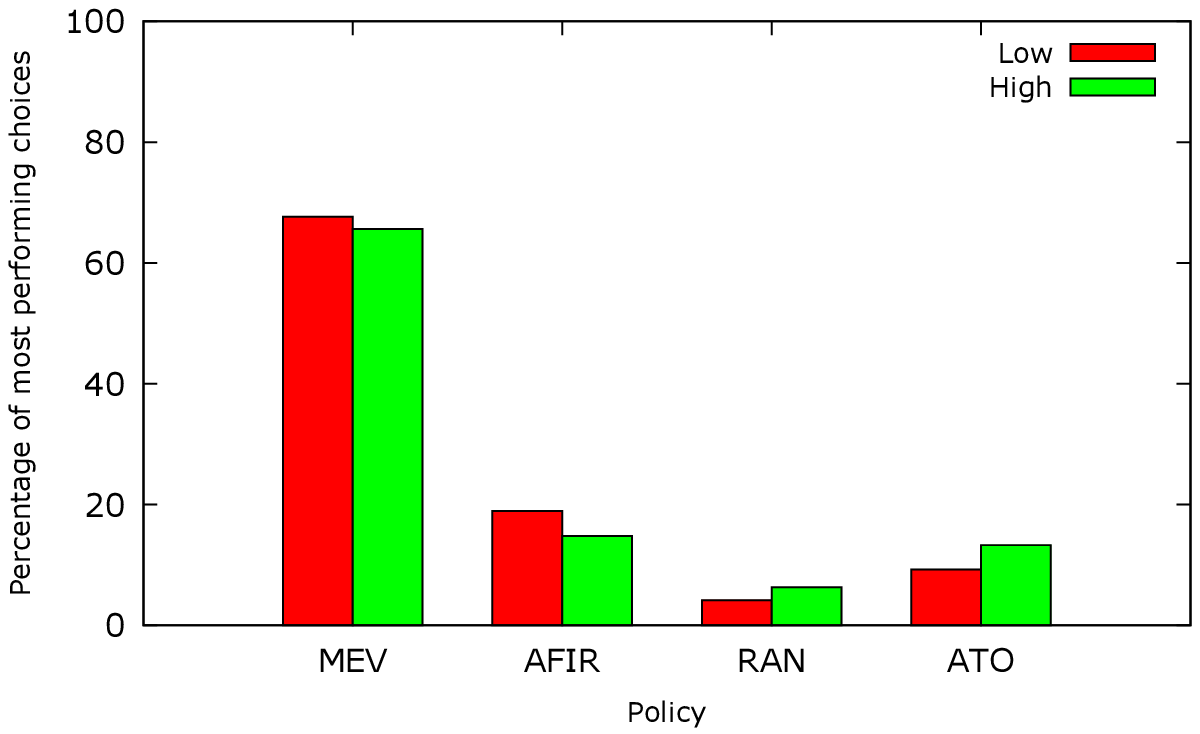}
	\vspace{-18pt}
	\caption{Percentage of selection of best composition w/o (left) and w/ (right) competing processes}
	\label{fig:percentages-best}
	\vspace{-8pt}
\end{figure}

Figure~\ref{fig:percentages-best} shows the percentage of time each policy identifies the best composition (i.e., the one with the minimum service provisioning time), varying the sizes of I/O parameters without CPU contention (left), and varying CPU contention (right). In the latter case, the size of I/O parameters is equal to 40KB. We anticipate that, in general, 40KB is the size of I/O where the advantage of MEV is lower, as MEV advantages become more and more evident as the network congestion increases. Therefore, Figure~\ref{fig:percentages-best} show ``how useful'' the model is in identifying the best composition among the considered policies.

Plots clearly show that MEV is by far the policy that selects more often the best composition. This holds true irrespective of the size of I/O parameters, or CPU contention. This happens in between 60\% and 70\% of the cases. AFIR is the next most successful policy, followed by ATO and RAN. We anticipate that this ranking among policies is a constant feature that we have found in our experiments. The fact that MEV outperforms AFIR shows the advantage of choosing service compositions based on an accurate model of the various phases of the service provisioning time. The fact that MEV outperforms ATO shows that composition is helpful as (i) it exploits service components that, due to mobility, may become available sooner to the seekers, and (ii) it spreads service load across providers more evenly. Finally, RAN yields the worst performance, as it does not exploit suitable service (component) providers upon encounters (as AFIR and MEV do), and uses composed services without evaluating whether this is advantageous over single-component services (as MEV does). 

\begin{figure}[htb]
	\centering
	\includegraphics[width=0.24\textwidth,height=0.2\textwidth]{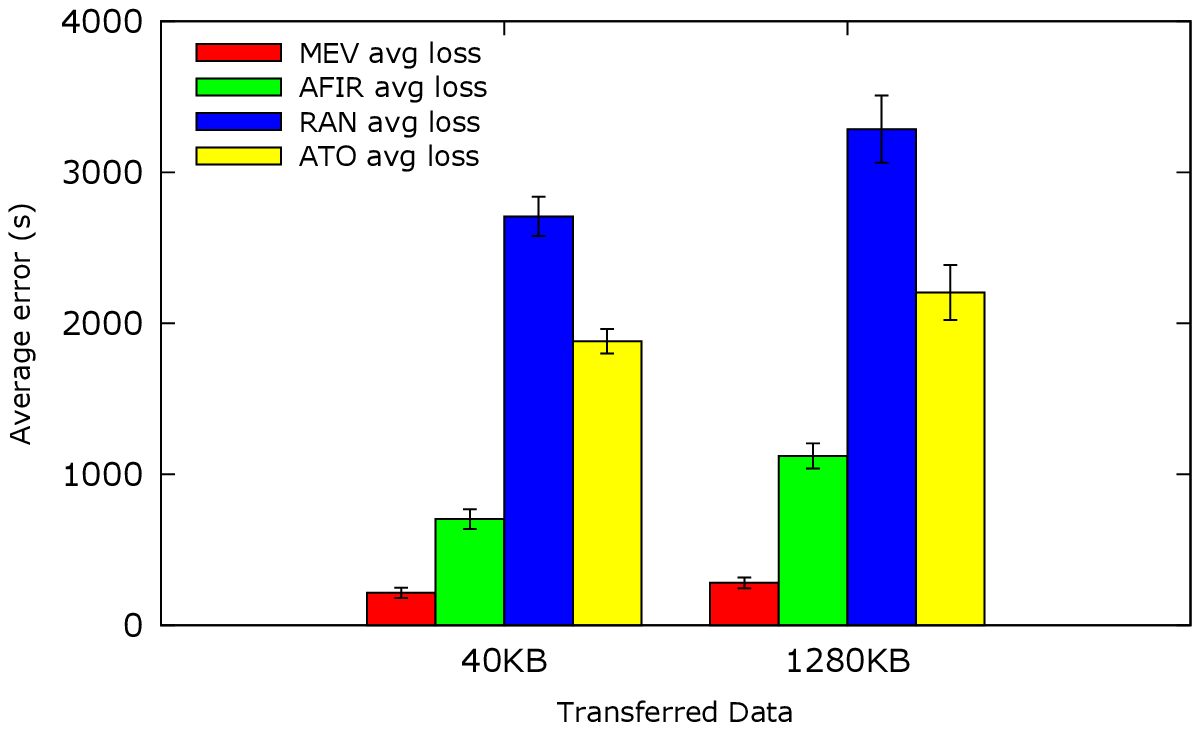}
	\includegraphics[width=0.24\textwidth,height=0.2\textwidth]{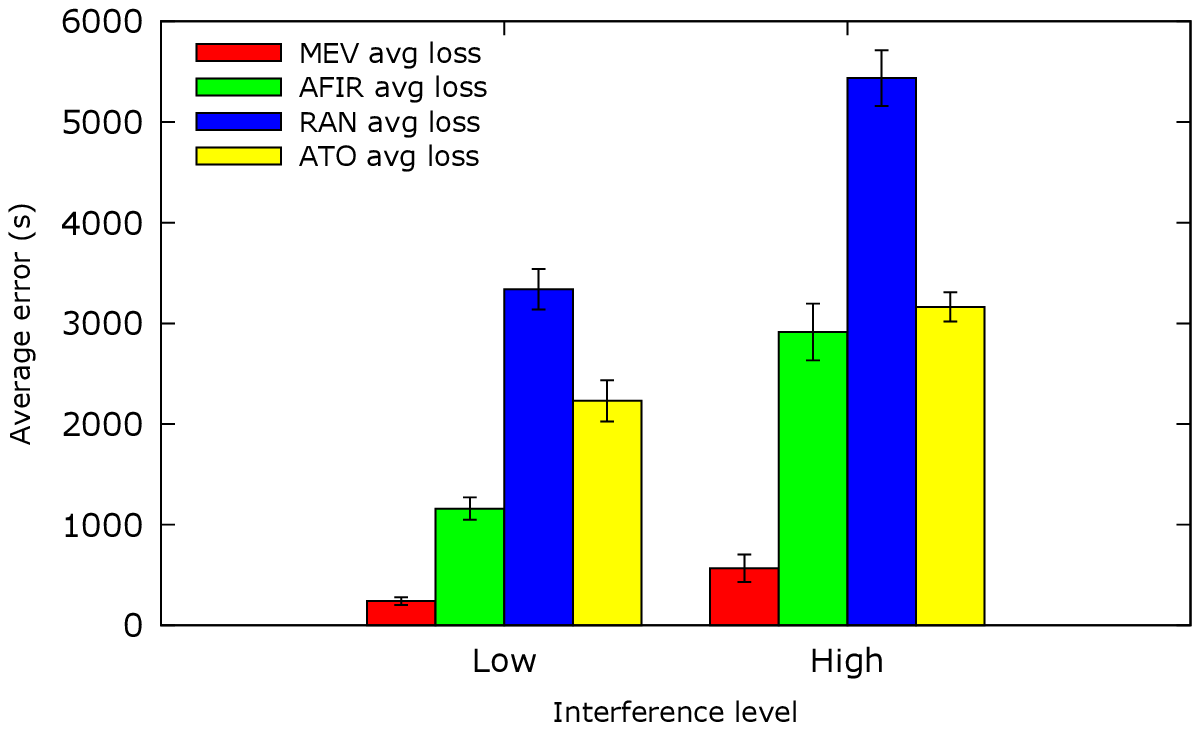}
	\vspace{-18pt}
	\caption{Average loss of the alternatives with respect to the best one, w/o (left) and w/ (right) competing processes}
	\label{fig:alternatives-loss}
	\vspace{-18pt}
\end{figure}

Figure~\ref{fig:alternatives-loss} shows the average ``loss'' in service provisioning time of each policy with respect to the best one, by measuring the additional delay in the service provisioning time with respect to the option which completes first for that service request. First of all, it is worth nothing that MEV is clearly the policy yielding the lowest average loss. Results confirms the same ranking among policies already observed in Figure~\ref{fig:percentages-best}. Finally, note that the advantage of MEV increases both with the size of I/O parameters, as well as with the number of competing processes, at least with respect to the second best, i.e., AFIR. This shows another general feature that we have found in our experiment, i.e., that the gain of MEV increases the more resources become limited, either in terms of bandwidth, or in terms of CPU at providers. The key reason is that our model takes into consideration the availability of both resources when estimating the expected service provisioning time, and is thus able to better distribute compositions among providers the more resources become scarce.

\subsubsection{Precision of model estimates}
\label{subsub:precision-estimates}

\begin{figure}[htb]
	\centering
	\includegraphics[width=0.24\textwidth,height=0.2\textwidth]{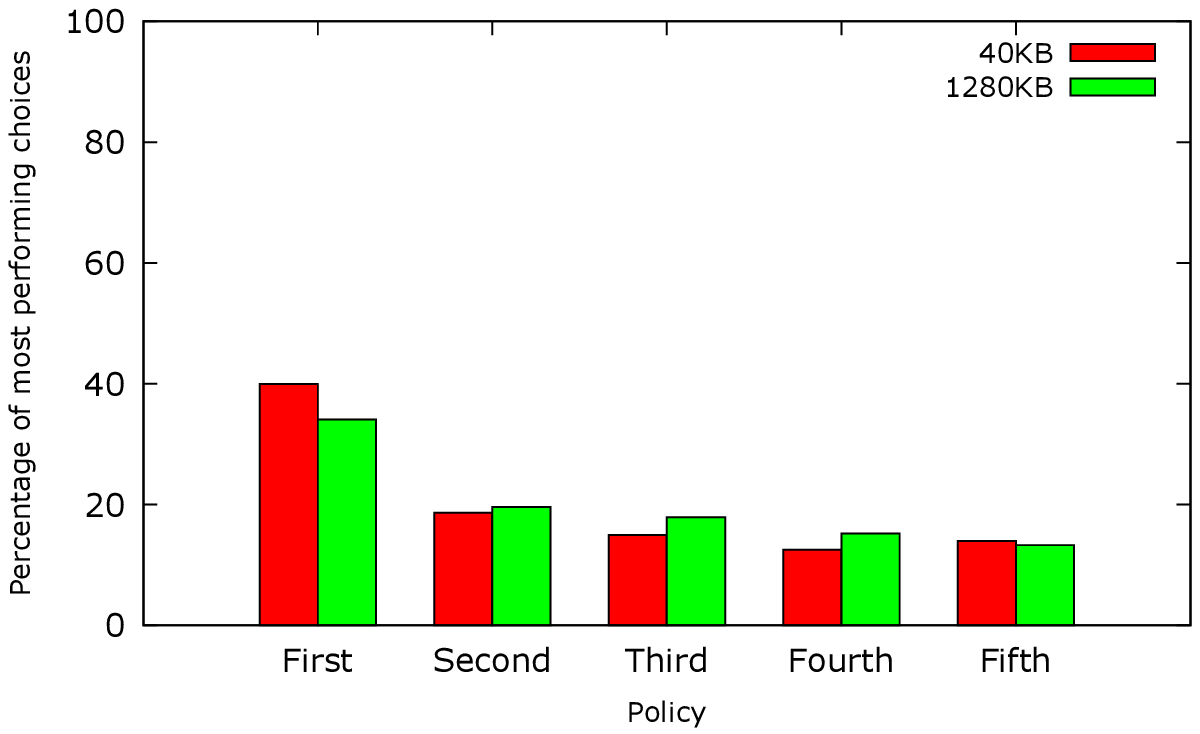}
	\includegraphics[width=0.24\textwidth,height=0.2\textwidth]{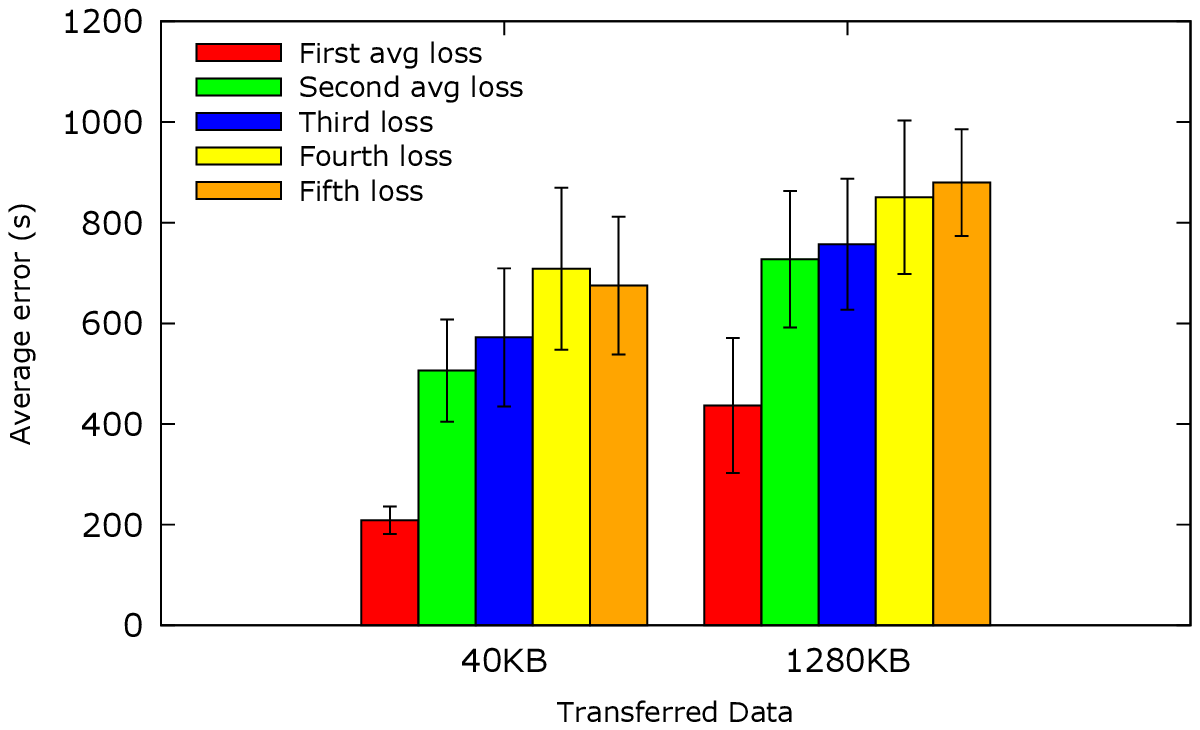}
	\vspace{-18pt}
	\caption{Percentage of ``success'' for the first five compositions (left), and average errors over the best (right)}
	\label{fig:MEV-alternatives}
	\vspace{-8pt}
\end{figure}

In this section, we analyse the effectiveness of the model not only in identifying the best composition, but, more broadly, in ranking alternatives. Specifically, we compare the first five service compositions ranked by our model (the first in the ranking being the one selected by MEV). Simulation settings and parameters are the same already explained in Section~\ref{subsub:precision-policies}. Figure~\ref{fig:MEV-alternatives} shows, on the left-hand side, the fraction of compositions for which each of the five alternatives (ranked by our model in increasing order of estimated service provisioning time) turns out being the best. The first choice of the model is the best composition in about 40\% of the cases (slightly less for 1280KB I/O sizes). This percentage is significantly higher than what a random selection among the five would achieve (20\%), and shows that the model is really helpful in ranking the alternatives. The fraction of times when the option ranked first by the model is best is significantly higher than the fraction when each of the other considered compositions is best. Interestingly, the model is able to rank the five alternatives rather precisely.  The fraction of times alternative $x$ in this ranking turns out being the best is normally higher than the percentage of time this happens for alternative $x+1$. Figure~\ref{fig:MEV-alternatives} indicates that there is room for improvement in MEV, as the model not always identifies the best composition, while ``second-best'' choices considered by the model result best in a significant fraction of the cases. While MEV could certainly be improved, results in Section~\ref{subsub:precision-policies} show that it is already much more precise than state-of-the art alternative policies. Based on Figure~\ref{fig:MEV-alternatives}, a possible improvement of MEV could be based on multi-path service composition: if MEV used, for example, the first three alternatives provided by the model and spawn three parallel executions of a service request, the fraction of times MEV would not use the best composition (among the five) would be only about 25\%. 

The plot in Figure~\ref{fig:MEV-alternatives} right-hand side shows, for each candidate composition provided by the model, the average difference in service provisioning time with the best. The ``loss'' increases when the network becomes more congested (higher I/O sizes), as in general service composition delays increase. However, in any case the loss of the first choice (which corresponds to MEV) is significantly lower than that of the other choices.

To complete the analysis, Table~\ref{tab:accuracy-estimates} shows the precision of the model in estimating the real average service composition time. Specifically, we show the average service composition time estimated by the model (``Est'') and the one measured in simulations for the compositions selected by MEV (``Sim''). We also include these results for the PMTR and Haggle-Info05 traces (differences in service provisioning times across traces clearly depend on the mobility patterns of the node in each one). Note that, our model estimates the \emph{average} service composition time that MEV would achieve,  while it does not provide an estimate of the distribution of the service provisioning time. In simulation, each sample of the MEV service composition time is a \emph{sample} from the random variable distribution whose average value our model estimates. Therefore, it would not be appropriate to compare the values of the model and the individual samples obtained in simulation, while the \emph{average} of those samples is the value to compare with the model. Table~\ref{tab:accuracy-estimates} shows  that the average service provisioning time predicted by the model is quite close to the one obtained in simulation for both traces and both data transfer sizes, with a maximum error of about 15\%. Remember that achieving a high precision on this index is not a primary goal of our model, whose main purpose is to rank alternatives. However, it is very good to see that the model is rather precise also from this standpoint. Note that, this also indicates that the model is robust in estimating the average service provisioning time with respect to the several simplifying assumptions we have to use in the analysis, as none of them holds in the simulations.

\begin{table}[htb]
	\vspace{-12pt}
	\caption{Accuracy of service provisioning time estimates}
	\vspace{-12pt}
	\centering
	\begin{tabular}{|p{0.3cm}|p{0.8cm}|p{0.9cm}|p{0.8cm}|p{0.9cm}|p{0.8cm}|p{0.9cm}|}
		\hline 
		& \multicolumn{2}{c|}{PMTR} & \multicolumn{2}{c|}{Info05}  &  \multicolumn{2}{c|}{RollerNet} \\ 
		\hline 
		& 40KB & 1280KB & 40KB & 1280KB & 40KB & 1280KB \\ 
		\hline 
		Est & $17413 \pm 1137$ & $13046 \pm 1305$ & $20138 \pm 1413$ & $21408 \pm1814$ & $474 \pm 58$ & $665 \pm 108$ \\ 
		\hline 
		Sim & $19091 \pm 2439$ & $15124 \pm 1750$ & $20208 \pm 1229$ & $19947 \pm 1624$ & $789 \pm 87$ & $916 \pm 196$ \\ 
		\hline 
	\end{tabular} 
\label{tab:accuracy-estimates}
\end{table}

Summarising, the results presented in this section clearly demonstrate the usefulness of the proposed model: (i) using the model, MEV is able to correctly rank the alternative options for service provisioning in a very large fraction of cases, and in particular to identify the best (among MEV, AFIR, RAN, ATO) in between 60\% and 70\% of the cases; (ii) MEV provides a clear gain over any other considered policy, by experiencing the lowest average loss with respect to the best; (iii) while the primary goal of the model is to rank service compositions, it is also quite precise in estimating the average service provisioning time for the composition chosen by MEV.

\subsection{Comparison between policies}
\label{sub:comparison-policies}

In this section we compare more directly MEV with AFIR, RAN and ATO (AFIR in greater detail, since it is consistently the second-best policy in the pool). We do so with respect to a number of parameters, i.e., different mobility traces, service execution times, size of I/O parameters, providers' CPU contention. For the case of MEV, we have also obtained results aimed at analysing more in detail the performance of the individual components of the service provisioning process, and the impact of longer or shorter warm-up periods (remember that the warm-up is needed to acquire reliable statistics about mobility, in order to feed the model). These results are presented in the Appendix and are omitted here, as they do not change any of the outcomes of the evaluation discussed in the following of this section.

In the following set of simulations, we slightly changed the configuration with respect to what explained in Section~\ref{subsub:precision-policies}. Specifically, we do not run all the four policies in parallel for each request. Rather, in each simulation we consider a specific policy, and draw the inter-request generation intervals from a uniform distribution in the range [40s,80s]. This setting allows us to guarantee that the request generation pattern is determined and controlled. The rest of the simulation settings are again as reported in Table~\ref{tab:ranking-sim-params}.

We first compare the four policies over all the mobility traces considered in the paper (PMTR, Haggle-Info05, Haggle-Info06, MIT-Reality, RollerNet). Note that these traces vary significantly in terms of number of nodes, patterns of mobility, contact and intercontact times. Specifically, they include scenarios where nodes move across large spaces and meet infrequently (e.g., in the case of Reality Mining) and scenarios where nodes move close-by and meet quite often (e.g., in the case of RollerNet). These features impact significantly on the reliability of intercontact time estimates used by MEV, and therefore the results presented allow us to characterise its performance over a wide range of mobility scenarios. Specifically, Table \ref{tab:PMTRPar} reports the simulation configurations used
with each of the traces.

\begin{table}[t]
	\caption {Trace-based simulation parameters}
	\centering{
		\scriptsize{
			\begin{tabular}{|p{1.55cm}|p{.95cm}|p{.95cm}|p{.95cm}|p{.95cm}|p{.95cm}|}
				\hline
				Parameters & PMTR & Info05 & Rollernet & Info06 & Reality\\
				\hline
				Number of nodes & $43$ & $55$ & $63$ & $50$ & $100$ \\
				\hline
				Total simulation time & $500000s$ & $250000s$ & $100000s$ & $330000s$ & $600000s$ \\
				\hline
				Mobility warm-up period & $50000s$ & $50000s$ & $30000s$ & $50000s$  & $20000s$ \\
				\hline
				Request generation phase duration& $450000s$ & $200000s$ & $70000s$ & $280000s$ & $580000s$ \\
				%\hline
				%Input/output data size & $40KB$ to $5120KB$\\
				\hline
				Density of each service & $75\%$  & $50\%$ & $75\%$ & $75\%$ & $75\%$\\
				\hline
			\end{tabular}
	}}
	\label{tab:PMTRPar}
	\vspace{-12pt}
\end{table}

Tables~\ref{tab:Time40KB} and \ref{tab:Time1280KB} show the average service provisioning time, together with their confidence intervals (with 95\% confidence level), for I/O sizes equal to 40KB and 1280KB, respectively. In the first row we can see the results of the simulations for PMTR traces. MEV yields 21\% and 43\% lower service provisioning time,  respectively, with respect to the second-best policy i.e., AFIR, and, as observed before, outperforms all the other policies too.  For the Info05 traces, MEV outperforms AFIR (by 4\% and 7\% respectively), and drastically outperforms the other policies.  MIT-Reality is the most challenging scenario for MEV (and the most favourable for AFIR). In fact, Reality is an extremely sparse trace, where the large variability of intercontact times and the low number of contact events make predictions not very reliable. However, even in such case, MEV and AFIR achieve comparable performance. The cases of Info06 and RollerNet exemplify mobility patterns over smaller areas, with shorter and more frequent inter-contact times. In the Info06 trace, MEV outperforms AFIR by 22\% and 37\%, for parameters sizes equal to 40KB and 1280KB, respectively. In the case of Rollernet, MEV outperforms AFIR by 21\% and 41\% in the two configurations. As before, RAN and ATO yield worse performance than AFIR (and MEV), with ATO being the best among the two. These results confirm that MEV outperforms AFIR (and all the other policies) by a very significant margin in a range of different mobility patterns. It is by large the best policy in scenarios with more "compact" mobility patterns. However, also in sparser conditions none of the alternative policies yield shorter service provisioning times.

\begin{table}[t]
	\caption {Average service provisioning time (s) - 40KB Scenario}
	\vspace{-12pt}
	\centering{
		\scriptsize{
			\begin{tabular}{|p{1.15cm}|p{1.3cm}|p{1.3cm}|p{1.3cm}|p{1.3cm}|}
				\hline
				& MEV & AFIR & RAN & ATO \\
				\hline
				PMTR & $35087 \pm 2209 $ & $44557 \pm 1897 $ & $65192 \pm 2929$ & $57020 \pm 2391$ \\
				\hline
				Info05 & $32828 \pm 1938$ & $34221 \pm 1057$ & $61614 \pm 3030$ & $49057 \pm 813 $  \\
				\hline
				Reality & $163358 \pm 6595$ & $161794 \pm 6611$ & $205849 \pm 15587$ & $197915 \pm 2697$  \\
				\hline
				Info06 & $15542 \pm 1718$ & $19999 \pm 1472$ & $56316 \pm 2035$ & $38298 \pm 2348$  \\
				\hline
				RollerNet & $788 \pm 86$  & $1007 \pm 65$ & $3533 \pm 142$ & $2110 \pm 97$\\
				\hline
			\end{tabular}
	}}
	\label{tab:Time40KB}
	\vspace{-8pt}
\end{table}

\begin{table}[t]
	\caption {Average service provisioning time (s) - 1280KB Scenario}
	\vspace{-12pt}
	\centering{
		\scriptsize{
			\begin{tabular}{|p{1.15cm}|p{1.3cm}|p{1.3cm}|p{1.3cm}|p{1.3cm}|}
				\hline
				& MEV & AFIR & RAN & ATO \\
				\hline
				PMTR & $26368 \pm 1452 $ & $46826 \pm 1179 $ & $66946 \pm 2223$ & $58834 \pm 1385$ \\
				\hline
				Info05 & $33835 \pm 513$ & $36568 \pm 1053$ & $62067 \pm 2559$ & $49841 \pm 1787 $  \\
				\hline
				Reality & $163105 \pm 7038$ & $158918 \pm 2722$ & $220052 \pm 4357$ & $197365 \pm 4999$  \\
				\hline
				Info06 & $13339 \pm 976$ & $21495 \pm 1606$ & $55954 \pm 3229$ & $37624 \pm 758$  \\
				\hline
				RollerNet & $915 \pm 195$  & $1562 \pm 177$ & $4293 \pm 126$ & $2573 \pm 275$\\
				\hline
			\end{tabular}
	}}
	\label{tab:Time1280KB}
	\vspace{-8pt}
\end{table}

\begin{figure}[t]
	\centering
	\includegraphics[width=0.4\textwidth]{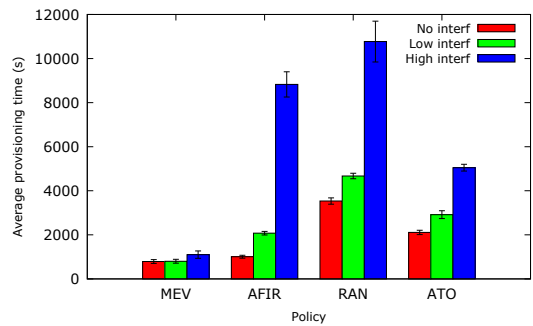}
	\caption{Service provisioning with changing CPU contention (RollerNet).}
	\vspace{-12pt}
	\label{fig:Roller_cpu}
\end{figure}

In Figure~\ref{fig:Roller_cpu} we analyse the behaviour of the four policies for different levels of CPU contention at providers. Together with results in Tables~\ref{tab:Time40KB} and \ref{tab:Time1280KB}, they assess the performance of the four policies under providers' resource constraints. The level of contention is controlled as explained at the beginning of Section~\ref{sec:results}. In particular, in the case of ``Low'' and ``High'' interference we have on average 3 and 10 other processes competing for the CPU at each provider. Thanks to the use of the model, which dynamically takes into consideration expected CPU load at providers, the performance of MEV is rather insensitive to the increase of CPU contention. This is achieved by distributing the load on less congested providers, as appropriate. It is quite interesting to note, among the other policies, the fact that AFIR suffers particularly under high contention, and the ranking between AFIR and ATO is swapped in this case. This is a side effect of AFIR generating longer compositions than ATO, which results in a higher total number of service component requests in the system. Under severe CPU contention, this is clearly a significant problem. Note that MEV is not affected by the same issue, as the model considers and compares compositions of any length, thus moving from longer to shorter compositions as appropriate.

\begin{figure}[t]
	\centering
	\includegraphics[width=0.4\textwidth]{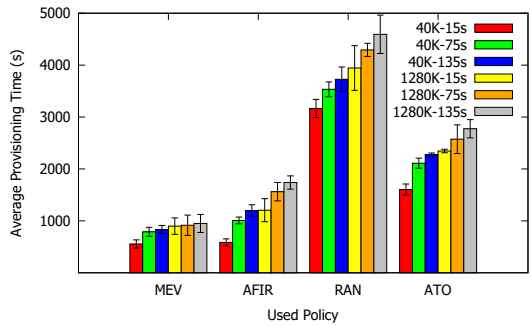}
	\caption{Service provisioning with changing service times (RollerNet).}
	\vspace{-12pt}
	\label{fig:Roller_service}
\end{figure}

The next set of results are obtained by varying the average service execution time at providers. To the best of our knowledge, no well-established benchmarks are available (such as a set of reference traces) to describe typical service execution times of personal mobile devices. Therefore, we have varied the average service execution time in our simulations, considering short (15s), medium (75s), and large (135s) average execution times. We believe these values are reasonable and representative, considering that we target services to be executed by personal mobile devices.  Therefore, we think that  the presented results allow us to characterise the performance of MEV in realistic conditions, also as far as service execution times are concerned. Note that, we have used the RollerNet trace, as this is the one with the shorter intercontact times, and therefore the one where service provisioning depends more on the service execution time. Figure \ref{fig:Roller_service} shows the service provisioning time for both the reference sizes of I/O parameters. In general, we observe that, as expected, the service provisioning time increases with the service execution time (and with the input/output sizes). However, note that in MEV the increase is very graceful, confirming that MEV is able to efficiently spread the load across the possible providers. While, in principle, also RAN would do the same, the comparison of performance clearly shows that using additional knowledge to estimate service provisioning times, as MEV does, is very useful. Also note that, the performance gap between AFIR and MEV constantly increases as service execution takes longer and longer.

\begin{figure}[htb]
	\centering
	\includegraphics[width=0.24\textwidth,height=0.2\textwidth]{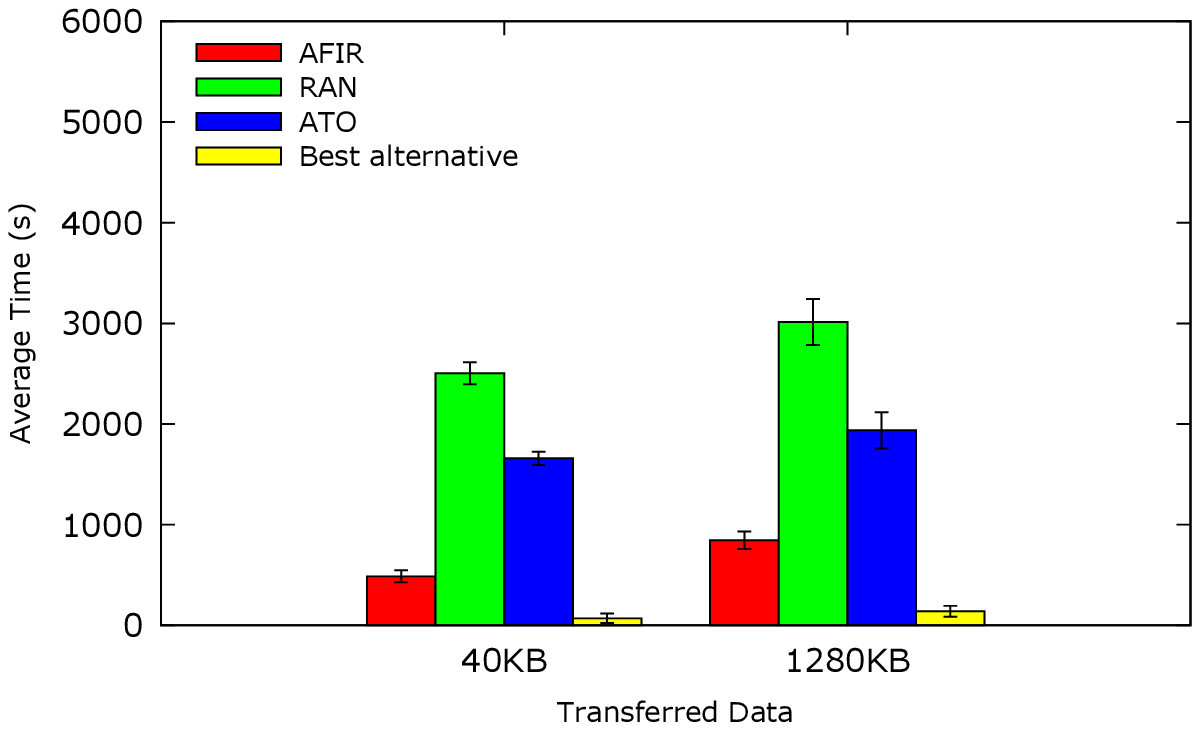}
	\includegraphics[width=0.24\textwidth,height=0.2\textwidth]{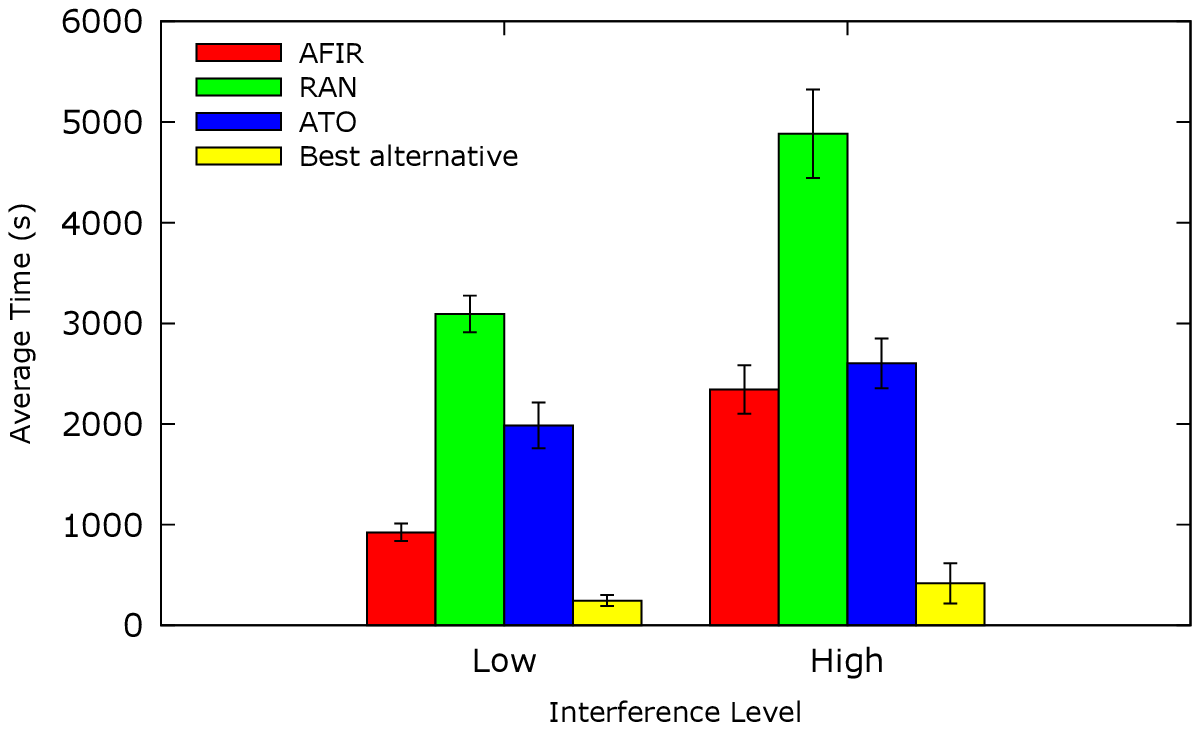}
	\vspace{-18pt}
	\caption{Average loss w/o (left) and w/ (right) competing processes}
	\label{fig:MEV-choices}
	\vspace{-8pt}
\end{figure}

The results presented so far clearly show an advantage of MEV over the other three policies. To complete the analysis, we show in Figure~\ref{fig:MEV-choices} the average ``loss'' (i.e., the average difference in service provisioning time) of the other policies against MEV. For this set of experiments, we used again the simulation settings explained in Section~\ref{subsub:precision-policies} to make sure time differences are computed for exactly the same requests, executed with each policy under the same network and providers' conditions. Results are shown for varying I/O sizes w/o CPU contention (left) and for varying levels of CPU contention with I/O sizes set to 40KB (right). The first three bars in the plot show the average loss of, respectively, AFIR, RAN and ATO with respect to MEV. The fourth bar shows the loss, with respect to MEV, of an ideal policy based on an oracle, which uses, among AFIR, RAN and ATO, the one that will perform best. Clearly, this policy is infeasible in practice. The plot shows that MEV provides a significant advantage in terms of service provisioning time over all of the three policies. Quite interestingly, MEV provides a non-negligible advantage over the ideal (infeasible) policy, too. Also in this case, these features do not qualitatively change with the size of I/O or the level of CPU contention. However, the advantage of MEV over the other policies increases when larger I/O sizes or higher contention are considered. This shows once more that, under shortage of resources (network bandwidth in the first case, CPU in the second case) MEV becomes more and more efficient.

\section{Conclusions} \label{Conclusions}
In this paper we have presented an approach (MEV) to identify effective service provisioning options in an opportunistic environment. We have defined a mathematical model by which seekers can estimate the expected service provisioning time using different available compositions, and thus select the best one. The model is based only on local knowledge that nodes collect by exchanging a few information between each other during contacts. We have analysed the \emph{precision} of the model in quite some detail, showing that the model fulfils its main objective, i.e., to identify better service provisioning options than the alternative policies we have considered. Specifically, in 60\% to 70\% of the cases, the model identifies a better service composition option with respect to the other policies. Its average ``loss'' with respect to the best policy is by far the lowest among the considered policies. The model is also able to correctly rank the first-$k$ compositions (we showed results with $k=5$). Specifically, ranking compositions according to the estimated service provisioning time of the model, the fraction of time the $i$-th composition turns out being the best one is higher than the fraction of time this happens for composition $i$+1 for all ranking positions $i$. 

We have then directly compared MEV with the alternative policies, AFIR, RAN, and ATO. MEV outperforms all of them by always yielding lower service provisioning times, across a range of very different realistic human mobility patterns. Specifically, we have found that MEV yields better performance for ``compact'' mobility patterns, i.e., when users do not move in too large areas, and meet rather frequently. Note that this is the case for the application cases we target and where we think opportunistic computing makes more sense. For example, in industrial applications, workers typically move in the area of a factory, and meet rather frequently over a day. In Mobile Social Networking applications, users are typically co-located in specific geographical areas (e.g., a museum, a theme park, etc.). Moreover, MEV outperforms the other policies over a range of parameters. Most notably, as the providers' network or computation resources becomes scarce, the gain of MEV increases, which shows that MEV is particularly suitable in resource-constrained environments. Finally, MEV even outperforms an ideal, infeasible, policy, which picks, for each request, the policy that will perform best for that request, among AFIR, RAN and ATO.

Taken together, these two sets of results show that (i) the model is able to correctly identify very good service compositions, and (ii) the gain of a policy using this model is very significant, across a range of significant parameters.
\vspace{-12pt}

\section*{Acknowledgments}
This work was partially funded by the EC under the H2020 REPLICATE (691735), SoBigData (654024) and AUTOWARE (723909) projects.

\vspace{-12pt}
\bibliographystyle{IEEEtran}

\begin{thebibliography}{10}
\providecommand{\url}[1]{#1}
\csname url@samestyle\endcsname
\providecommand{\newblock}{\relax}
\providecommand{\bibinfo}[2]{#2}
\providecommand{\BIBentrySTDinterwordspacing}{\spaceskip=0pt\relax}
\providecommand{\BIBentryALTinterwordstretchfactor}{4}
\providecommand{\BIBentryALTinterwordspacing}{\spaceskip=\fontdimen2\font plus
\BIBentryALTinterwordstretchfactor\fontdimen3\font minus
  \fontdimen4\font\relax}
\providecommand{\BIBforeignlanguage}[2]{{%
\expandafter\ifx\csname l@#1\endcsname\relax
\typeout{** WARNING: IEEEtran.bst: No hyphenation pattern has been}%
\typeout{** loaded for the language `#1'. Using the pattern for}%
\typeout{** the default language instead.}%
\else
\language=\csname l@#1\endcsname
\fi
#2}}
\providecommand{\BIBdecl}{\relax}
\BIBdecl

\bibitem{cisco2016}
Cisco. (2016, February) Cisco visual networking index: Global mobile data
  traffic forecast update, 2015 - 2020.

\bibitem{Vincentelli:2015aa}
A.~S. Vincentelli, ``Let's get physical: Adding physical dimensions to cyber
  systems,'' in \emph{Low Power Electronics and Design (ISLPED), 2015 IEEE/ACM
  International Symposium on}, July 2015, pp. 1--2.

\bibitem{pppeu}
\BIBentryALTinterwordspacing
5G-PPP. White paper on factories-of-the-future vertical sector. [Online].
  Available: \url{https://5g-ppp.eu/white-papers/}
\BIBentrySTDinterwordspacing

\bibitem{GarciaLopez:2015:ECV:2831347.2831354}
P.~Garcia~Lopez, A.~Montresor, D.~Epema, A.~Datta, T.~Higashino, A.~Iamnitchi,
  M.~Barcellos, P.~Felber, and E.~Riviere, ``Edge-centric computing: Vision and
  challenges,'' \emph{ACM CCR}, vol.~45, no.~5, pp. 37--42, Sep. 2015.

\bibitem{Habak}
K.~Habak, M.~Ammar, K.~A. Harras, and E.~Zegura, ``{FemtoClouds: Leveraging
  Mobile Devices to Provide Cloud Service at the Edge},'' in \emph{IEEE 8th
  Int. Conf. Cloud Comput.}, 2015.

\bibitem{sap2017}
\BIBentryALTinterwordspacing
SAP. (2017, January) Forrester report: Connect the edge and core to power your
  business. [Online]. Available:
  \url{https://www.sap.com/cmp/dg/connect-edge-core/index.html}
\BIBentrySTDinterwordspacing

\bibitem{Conti2013}
M.~Conti, E.~Marzini, D.~Mascitti, A.~Passarella, and L.~Ricci, ``Service
  selection and composition in opportunistic networks,'' in \emph{Proc. IWCMC},
  July 2013, pp. 1565--1572.

\bibitem{Conti2010}
M.~Conti and M.~Kumar, ``Opportunities in opportunistic computing,'' \emph{IEEE
  Computer}, vol.~43, no.~1, pp. 42--50, Jan 2010.

\bibitem{3GPP2016}
3GPP, ``Proximity-based services (prose); stage 2,'' May 2016,
  http://www.3gpp.org/DynaReport/23303.htm.

\bibitem{Keraenen2009}
A.~Ker\"{a}nen, J.~Ott, and T.~K\"{a}rkk\"{a}inen, ``The {ONE} simulator for
  dtn protocol evaluation,'' in \emph{Proc. Simutools}, 2009, pp. 55:1--55:10.

\bibitem{Kalasapur2007}
S.~Kalasapur, M.~Kumar, and B.~Shirazi, ``Dynamic service composition in
  pervasive computing,'' \emph{IEEE TPDS}, vol.~18, no.~7, pp. 907--918, July
  2007.

\bibitem{Bianchini2005}
D.~Bianchini, V.~De~Antonellis, and M.~Melchiori, ``An ontology-based
  architecture for service discovery and advice system,'' in \emph{Proc. DEXA},
  Aug 2005, pp. 551--556.

\bibitem{DelPrete2008}
L.~{Del Prete} and L.~Capra, ``Reliable discovery and selection of composite
  services in mobile environments,'' in \emph{In Proc. of 12th IEEE EDOC},
  2008.

\bibitem{Wang2011}
J.~Wang, ``Exploiting mobility prediction for dependable service composition in
  wireless mobile ad hoc networks,'' \emph{IEEE Trans. Serv. Comput.}, vol.~4,
  no.~1, pp. 44--55, Jan. 2011.

\bibitem{Huerta-Canepa2010}
G.~Huerta-Canepa and D.~Lee, ``A virtual cloud computing provider for mobile
  devices,'' in \emph{Proc. ACM MCC}, 2010.

\bibitem{Tamhane2012}
S.~Tamhane, M.~Kumar, A.~Passarella, and M.~Conti, ``Service composition in
  opportunistic networks,'' in \emph{Proc. IEEE CPSCOM}, Nov 2012.

\bibitem{Sadiq2015}
U.~Sadiq, M.~Kumar, A.~Passarella, and M.~Conti, ``Service composition in
  opportunistic networks: A load and mobility aware solution,'' \emph{IEEE
  ToC}, vol.~64, no.~8, pp. 2308--2322, Aug 2015.

\bibitem{Passarella2011}
A.~Passarella, M.~Kumar, M.~Conti, and E.~Borgia, ``Minimum-delay service
  provisioning in opportunistic networks,'' \emph{IEEE TPDS}, vol.~22, no.~8,
  pp. 1267--1275, Aug 2011.

\bibitem{Shi2012_SER}
C.~Shi, V.~Lakafosis, M.~H. Ammar, and E.~W. Zegura, ``Serendipity: Enabling
  remote computing among intermittently connected mobile devices,'' in
  \emph{Proc. ACM MobiHoc}, 2012.

\bibitem{groba2014}
C.~Groba and S.~Clarke, ``Opportunistic service composition in dynamic ad hoc
  environments,'' \emph{IEEE Trans. Serv. Comp.}, vol.~7, no.~4, pp. 642--653,
  Oct 2014.

\bibitem{Shi2012}
C.~Shi, M.~H. Ammar, E.~W. Zegura, and M.~Naik, ``Computing in cirrus clouds:
  The challenge of intermittent connectivity,'' in \emph{ACM MCC}, 2012.

\bibitem{Mtibaa2013a}
A.~Mtibaa, A.~Fahim, K.~Harras, and M.~Ammar, ``{Towards resource sharing in
  mobile device clouds: Power balancing across mobile devices},'' \emph{Proc.
  ACM MCC}, 2013.

\bibitem{Mtibaa2013}
A.~Mtibaa, K.~a. Harras, and A.~Fahim, ``{Towards Computational Offloading in
  Mobile Device Clouds},'' \emph{2013 IEEE 5th Int. Conf. Cloud Comput.
  Technol. Sci.}, 2013.

\bibitem{Mtibaa}
A.~Mtibaa, K.~A. Harras, K.~Habak, M.~Ammar, and E.~W. Zegura, ``{Towards
  Mobile Opportunistic Computing},'' in \emph{IEEE 8th Int. Conf. Cloud
  Comput.}, 2015.

\bibitem{Erramilli2008}
V.~Erramilli, M.~Crovella, A.~Chaintreau, and C.~Diot, ``Delegation
  forwarding,'' in \emph{Proc. ACM MobiHoc}, 2008.

\bibitem{Mascitti2014}
D.~Mascitti, M.~Conti, A.~Passarella, and L.~Ricci, ``Service provisioning
  through opportunistic computing in mobile clouds,'' \emph{Procedia Computer
  Science}, vol.~40, no.~0, pp. 143 -- 150, 2014.

\bibitem{Gao2009}
W.~Gao, Q.~Li, B.~Zhao, and G.~Cao, ``Multicasting in delay tolerant networks:
  A social network perspective,'' in \emph{Proc. ACM MobiHoc}, 2009.

\bibitem{Tournoux2011}
P.~Tournoux, J.~Leguay, F.~Benbadis, J.~Whitbeck, V.~Conan, and M.~Dias~de
  Amorim, ``Density-aware routing in highly dynamic dtns: The rollernet case,''
  \emph{IEEE TMC}, vol.~10, no.~12, pp. 1755--1768, Dec 2011.

\bibitem{Spyropoulos2009}
T.~Spyropoulos, T.~Turletti, and K.~Obraczka, ``Routing in delay-tolerant
  networks comprising heterogeneous node populations,'' \emph{IEEE TMC},
  vol.~8, no.~8, pp. 1132--1147, Aug 2009.

\bibitem{Takagi1991}
H.~Takagi, \emph{Vacation and Priority Systems, Part 1}.\hskip 1em plus 0.5em
  minus 0.4em\relax Elsevier Science Publishers B.V., 1991.

\bibitem{pmtr-2008}
P.~Meroni, S.~Gaito, E.~Pagani, and G.~P. Rossi, ``{CRAWDAD} dataset unimi/pmtr
  (v. 2008-12-01),'' Dec. 2008.

\bibitem{haggle-2009}
J.~Scott, R.~Gass, J.~Crowcroft, P.~Hui, C.~Diot, and A.~Chaintreau,
  ``{CRAWDAD} dataset cambridge/haggle (v. 2009-05-29),'' May 2009.

\bibitem{mit-reality-2005}
N.~Eagle and A.~S. Pentland, ``{CRAWDAD} dataset mit/reality (v. 2005-07-01),''
  Jul. 2005.

\bibitem{rollernet-2009}
F.~Benbadis and J.~Leguay, ``{CRAWDAD} dataset upmc/rollernet (v.
  2009-02-02),'' Feb. 2009.

\bibitem{Navidi2004}
W.~Navidi and T.~Camp, ``Stationary distributions for the random waypoint
  mobility model,'' \emph{IEEE TMC}, vol.~3, no.~1, pp. 99--108, Jan 2004.

\bibitem{Boldrini2010}
C.~Boldrini and A.~Passarella, ``Hcmm: Modelling spatial and temporal
  properties of human mobility driven by users' social relationships,''
  \emph{Comput. Commun.}, vol.~33, no.~9, pp. 1056--1074, Jun. 2010.

\end{thebibliography}

\vspace{-40pt}
\begin{IEEEbiography}[{\includegraphics[width=1in,height=1.25in,clip,keepaspectratio]{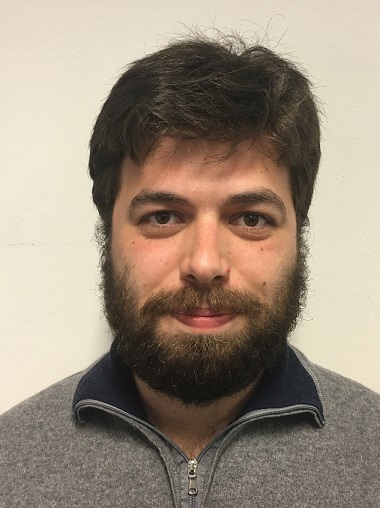}}]{Davide Mascitti}
	is an R\&D Cloud Developer at BioBeats. Prior to that, he was a research fellow at the Institute for Informatics and Telematics of the Italian National Research Council. Davide holds a PhD in Computer Engineering from the University of Pisa. His research focuses on designing and modeling self-organizing mobile systems and on cloud-assisted mobile computing.
\end{IEEEbiography}

\vspace{-40pt}
\begin{IEEEbiography}[{\includegraphics[width=1in,height=1.25in,clip,keepaspectratio]{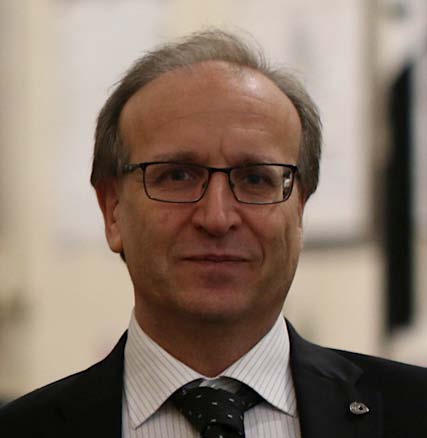}}]{Marco Conti}
is a research director and scientific counselor of the Italian National Research Council.  He has published more than 400 scientific papers related to design, modelling, and experimentation of Internet architecture and protocols, pervasive systems and social networks. He is the founding EiC of Online Social Networks and Media, EiC of Computer Communications and EiC for special issues of Pervasive and Mobile Computing. He has received several awards, including the Best Paper Award at IFIP Networking 2011, IEEE ISCC 2012 and IEEE WoWMoM 2013. He served as TPC/General chair for several major conferences, including IFIP Networking 2002, IEEE WoWMoM 2005 and 2006, IEEE PerCom 2006 and 2010, and ACM MobiHoc 2006.
\end{IEEEbiography}

\vspace{-40pt}
\begin{IEEEbiography}[{\includegraphics[width=1in,height=1.25in,clip,keepaspectratio]{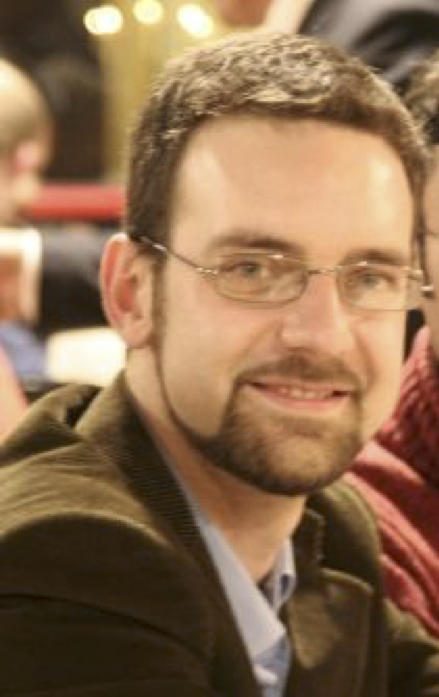}}]{Andrea Passarella}
(PhD Comp. Eng. 05) is with IIT-CNR, Italy. He was a Research Associate at the Computer Laboratory, Cambridge, UK. He published 140+ papers on mobile social networks, opportunistic, mobile edge computing, pervasive networks, receiving best paper awards at, among others, IFIP Networking 2011 and IEEE WoWMoM 2013. He co-authored the book ``Online Social Networks'' (Elsevier, 2015). He was PC Co-Chair of IEEE WoWMoM 2011, Workshops Co-Chair of IEEE PerCom and WoWMom 2010, ACM MobiSys 2015, and Co-Chair of several IEEE and ACM workshops. He is in the Associate EiC of Elsevier OSNEM, and Area Editor for Elsevier Pervasive and Mobile Computing. He is the Chair of the IFIP TC6 WG 6.3 Performance of Communication Systems.
\end{IEEEbiography}

\vspace{-40pt}
\begin{IEEEbiography}[{\includegraphics[width=1in,height=1.25in,clip,keepaspectratio]{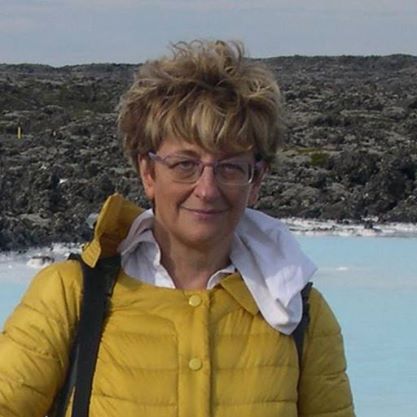}}]{Laura Ricci}
Laura Ricci is an Associate Professor of the Department of Computer Science, University of Pisa where she teaches several courses in the area of Computer Networks. Her main research interests are in the field of distributed computing, in particular P2P, blockchains and data intensive computing. She is the co-chair of the Large Scale Distributed Virtual Environments, LSDVE, workshop series, held every year in conjunction with EUROPAR. She has been the guest editor of several special issues in international journals and has chaired several workshops in International Conferences. Laura Ricci is author of more than 120 papers published in in refereed journals, books and international conference proceedings.
\end{IEEEbiography}

\vspace{-40pt}
\begin{IEEEbiography}[{\includegraphics[width=1in,height=1.25in,clip,keepaspectratio]{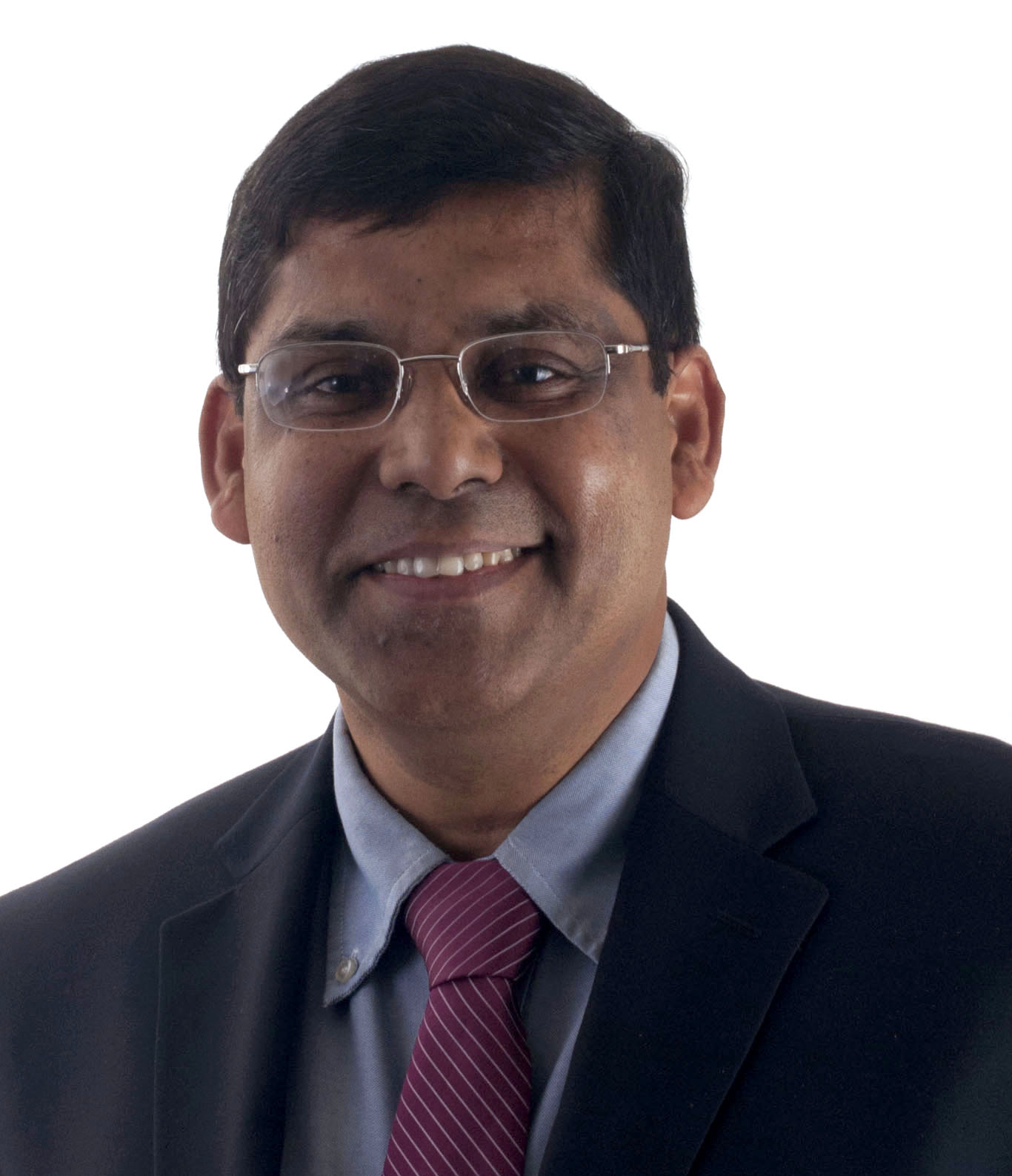}}]{Sajal K. Das}
(F'15) is a professor and Daniel St. Clair Endowed Chair of Computer Science at Missouri University of Science and Technology. His research interests include wireless sensor networks, cyber physical systems and IoT, smart environments (smart city, smart grid and smart health care), cloud computing, cyber security, and social networks. He has published over 700 papers, holds 5 US patents and coauthored 4 books. His h-index is 79 with more than 26,000 citations. He is a recipient of 10 Best Paper Awards at conferences like ACM MobiCom and IEEE PerCom, and IEEE Computer Society’s Technical Achievement Award for pioneering contributions to sensor networks and mobile computing. He serves as founding Editor-in-Chief of Pervasive and Mobile Computing Journal, and as Associate Editor of IEEE Transactions of Mobile Computing and ACM Transactions on Sensor Networks. 
\end{IEEEbiography}

\newpage
\clearpage
\appendices
\section{}
\subsection{Proofs of Lemmas for data transfer times in Case 2}
\label{appendix}
\begin{lemma}
\label{AlemN2A}
The probability that, in case 2A, $\theta$ includes $N_{2A}=n$ additional inter-contact times (after the first one that it always includes) is as follows:
\begin{equation}
P\{N_{2A}=n\}=e^{-\delta\frac{k+k'}{V}}*\frac{(\frac{\delta k'}{V})^{n+1}}{n+1!}*\frac{(1-\rho)\mu}{\delta+\mu(1-\rho)}
\end{equation}
\end{lemma}
\begin{IEEEproof}
We denote by $T_{CR}(0)$ the part of the first contact time between the seeker and the provider, that is the initial part of $\theta$. Therefore, the remaining transfer time is $k'/V-T_{CR}(0)$. In order for $N_{2A}$ to be equal to $n$, the remaining transfer time must be longer than $n$ contact times but shorter than $n+1$ contact times. Therefore, 
\begin{multline}
P\{N_{2A}=n\}=\\
=P\{\sum_{i=1}^{n}T_{C}(i) < \frac{k'}{V} - T_{CR}(0) < \sum_{i=1}^{n+1}T_{C}(i) | T_{CR}(0)<\frac{k'}{V}\}
\end{multline}
Given that all contact periods $T_{C}(i)$ between $h$ and $j$ are independent and identically distributed exponential random variables, we can consider $\sum_{i=1}^{n} T_{C}(i)$ as an Erlang variable $S_{C,n}$ with $n$ components, each with rate $\delta$.

In Lemma \ref{AlemTCR0} (see the Appendix) we derive the distribution of  $T_{CR}(0)$. Based on this result, we can thus condition on a specific duration of  $T_{CR}(0)$ and apply the law of total probability:
\begin{multline*}
P\{N_{2A}=n\}=\\
=\int_0^{\frac{k'}{V}}P\{S_{C,n}<\frac{k'}{V}-t<S_{C,n}+T_{C}(n+1)|\\
| T_{CR}(0)=t\}*P\{T_{CR}(0)=t\}dt=
\end{multline*}

\begin{multline*}
=\int_0^{\frac{k'}{V}}\int_0^{\frac{k'}{V}-t}\left(1-F_{T_{C}(n+1)}\left(\frac{k'}{V}-t-x\right)\right)*\\
\\*f_{S_{C,n}}(x)*f_{T_{CR}(0)}(t)dx dt=
\end{multline*}
Recalling that $T_{C}(n+1)$ is exponentially distributed, while $S_{C,n}$ follows an Erlang distribution, we obtain 
\begin{equation}
\label{eqn:cumulc}
F_{T_{C}(n+1)}(t)=1-e^{-\delta t}
\end{equation}
and
\begin{equation}
\label{eqn:erlang}
f_{S_{C,n}}(t)=\frac{\delta^{n}t^{n-1}e^{-\delta t}}{(n-1)!}
\end{equation}
Therefore, we obtain:
\begin{multline}
P\{N_{2A}=n\}=\\
=\int_0^{\frac{k'}{V}}\int_0^{\frac{k'}{V}-t}e^{-\delta(\frac{k'}{V}-t-x)}*\frac{\delta^n x^{n-1} e^{-\delta x}}{n-1!}*\\
*\frac{\delta e^{-\delta(t+\frac{k}{V})}*(1-\rho) \mu}{\delta+\mu(1-\rho)} dx dt=\\
=e^{-\delta\frac{k+k'}{V}}*\frac{(\frac{\delta k'}{V})^{n+1}}{n+1!}*\frac{(1-\rho)\mu}{\delta+\mu(1-\rho)}
\end{multline}
\end{IEEEproof}

\begin{lemma}
\label{AlemTCR0}
The probability density function of the residual of the first contact starting after the service component execution in case 2A is equal to:
\begin{equation*}
f_{T_{CR}(0)|case2A}(t)=\frac{\delta e^{-\delta(t+\frac{k}{V})}*\mu(1-\rho)}{\delta+\mu(1-\rho)}\\
\end{equation*}
\end{lemma}
\begin{IEEEproof}
Finding this probability density means we have to find a formulation for:
\begin{multline*}
f_{T_{CR}(0)|case2A}(t)=P\{T_{CR}(0)=t|\\
|DQ_{j}+DS_{s_i,j}+\frac{k+k'}{V}>T_{C}(0)> DQ_{j}+DS_{s_i,j}+\frac{k}{V}\}\\
\end{multline*}
We substitute the residual with the total value of the duration of the first contact $T_{CR}$
\begin{multline*}
f_{T_{CR}(0)|case2A}(t)=\\
=P\{T_{C}(0)-(DQ_{j}+DS_{s_i,j}+\frac{k}{V})=t|\\
|DQ_{j}+DS_{s_i,j}+\frac{k+k'}{V}>T_{C}(0)> DQ_{j}+DS_{s_i,j}+\frac{k}{V}\}=
\end{multline*}
\begin{multline*}
=\int_0^{\infty}P\{T_{C}(0)-\left(x+\frac{k}{V}\right)=t|DQ_{j}+DS_{s_i,j}=x\wedge\\
\wedge x+\frac{k+k'}{V}>T_{C}(0)> x+\frac{k}{V}\}*f_{DQ_{j}+DS_{s_i,j}}(x)dx
\end{multline*}
To make the analysis tractable, we model the provider as an $M/M/1$ queue, and therefore the service execution times $DS_{s_i,j}$ can be considered exponentially distributed with the provider having average service rate $\mu$ and load $\rho$. Under this assumption, the probability density function $f_{DS_{s_i,j}+DQ_{j}}$ of $DS_{s_i,j}+DQ_{j}$ is:
\begin{equation}
\label{eqn:Aeserv}
f_{DS_{s_i,j}+DQ_{j}}(t)=(1-\rho) \mu e^{-\mu (1- \rho)t}
\end{equation}
With this value, we obtain:
\begin{multline*}
f_{T_{CR}(0)|case2A}(t)=\\
=\int_0^{\infty}f_{T_{C}(0)}\left(t+x+\frac{k}{V}\right)*(1-\rho)\mu e^{-\mu (1-\rho)x}dx=\\
=\int_0^{\infty}\delta e^{-\delta \left(t+x+\frac{k}{V}\right)}*(1-\rho)\mu e^{-\mu (1-\rho)x}dx
\end{multline*}
that through integration gives us the result:
\begin{equation}
f_{T_{CR}(0)|case2A}(t)=\frac{\delta e^{-\delta(t+\frac{k}{V})}*\mu(1-\rho)}{\delta+\mu (1-\rho)}
\end{equation}
\end{IEEEproof}

\begin{lemma}
The probability of the first contact lasting enough to complete the input data transfer, but not enough to reach the beginning of the output data transfer, is:
\begin{equation}
p(Case2A)=\frac{e^{-\delta\frac{k}{V}}*\mu(1-\rho)}{\delta+\mu(1-\rho)}
\end{equation}
\end{lemma}
\begin{IEEEproof}
This probability can be written as the probability that the first contact time $T_{C}(0)$ is longer than the time for the transfer of the input parameters ($k/V$) plus the queuing time at the provider ($DQ_j$) plus the service computation time ($DS_{i,j}$), but not long enough to also include the transfer of the output parameters ($\theta$). By recalling that we are in case 2 and, therefore, $T_{C}(0)$ is not shorter than $k/V$ (captured in case 3) and shorter than the total service provisioning time without any disconnection (case 1), we can write:
\begin{multline*}
p(Case2A)=P\{\frac{k}{V}+DQ_j+DS_{i,j}<T_{C}(0)|\\
|\frac{k}{V}<T_{C}(0)<\frac{k}{V}+DQ_j+DS_{i,j}+\frac{k'}{V}\}
\end{multline*}
That, isolating the durations of $DQ_j$ and $DS_{i,j}$ becomes:
\begin{multline*}
p(Case2A)=\int_0^{\infty}P\{\frac{k}{V}+x<T_{C}(0)|\\
|\frac{k}{V}<T_{C}(0)<\frac{k}{V}+x+\frac{k'}{V} \wedge DQ_j+DS_{i,j}=x\}*\\
	*P\{DQ_j+DS_{i,j}=x\}dx=\\
=\int_0^{\infty}(1-F_{T_{C}(0)}(\frac{k}{V}+x))*f_{DQ_j+DS_{i,j}}(x)dx
\end{multline*}
Using the expression for the density of $DQ_j+DS_{i,j}$ derived in Lemma \ref{AlemTCR0} and the formula for the cumulative probability of $T_{C}(0)$
\begin{multline*}
p(Case2A)=\\
=\int_0^{\infty}(1-(1-e^{-\delta(\frac{k}{V}+x)}))*(1-\rho)\mu e^{-\mu(1-\rho)x}dx=\\
%=\int_0^{\infty}e^{-\delta\frac{k}{V}}*e^{-\delta x}*(1-\rho)\mu e^{-\mu(1-\rho)x}dx=\\
=e^{-\delta\frac{k}{V}}\mu(1-\rho)\int_0^{\infty}e^{-x(\delta+\mu(1-\rho))}dx=\frac{e^{-\delta\frac{k}{V}}*\mu(1-\rho)}{\delta+\mu(1-\rho)}
\end{multline*}
\end{IEEEproof}

\begin{lemma}
\label{AlemN}
The probability that, in cases 2B and 2C, 
theta includes $N_{2B}=n$ inter-contact times is as follows:
\begin{multline}
\label{Aenic}
P\{N_{2B}=n\}=P\{N_{2C}=n\}=\\
=P\{\sum_{i=1}^{n} T_{C}(i) < \frac{k}{V} < \sum_{i=1}^{n+1} T_{C}(i)\}=\\
=e^{-\delta \frac{k}{V}}*\frac{(\frac{\delta k}{V})^n}{n!}
\end{multline}
\end{lemma}
\begin{IEEEproof}
Analogously to Lemma \ref{lemN2A}, we can use the Erlang random variable $S_{C,n}$ to represent the first $n$ contacts of the data transfer phase , obtaining:
\begin{multline}
P\{N_{2B}=n\}=P\{N_{2C}=n\}=\\
= P\{S_{C,n} < \frac{k}{V} < S_{C,n}+  T_{C}(n+1)\}
\end{multline}
To find this value, we rewrite the expression in order to use the known formulas for the density function of Erlang distributions and the cumulative probability function for exponential distributions seen in Equation \ref{eqn:cumulc} and Equation \ref{eqn:erlang}:

\begin{multline}
P\{S_{C,n} < \frac{k}{V} < S_{C,n}+  T_{C}(n+1)\}=\\
 =\int_0^\frac{k}{V}P\{\frac{k}{V} - x < T_{C}(n+1) | S_{C,n}=x\}*f_{S_{C,n}}(x) dx=\\
 =\int_0^\frac{k}{V}(1-F_{T_{C}(n+1)}(\frac{k}{V}-x))*f_{S_{C,n}}(x)dx=\\
=\int_0^\frac{k}{V} e^{-\delta(\frac{k}{V}-x)}*\frac{\delta^{n}x^{n-1}e^{-\delta x}}{(n-1)!}dx=\\
=\frac{ e^{-\delta\frac{k}{V}} \delta^{n+1} }{n-1!}\int_0^{\frac{k}{V}} x^{n-1} dx =\frac{ e^{-\delta\frac{k}{V}}\delta^{n+1} }{n-1!}* \frac{(\frac{k}{V})^n}{n}=\\
=e^{-\delta \frac{k}{V}}*\frac{(\frac{\delta k}{V})^n}{n!}\\
\end{multline}
that is the result provided in the lemma.
\end{IEEEproof}

\begin{lemma}
\label{AlemcaseA}
The expected value of $\theta| case2A$ is equal to: 
\begin{multline}
E[\theta| case2A]=\frac{k'}{V}+\frac{1}{\delta'}+\frac{1}{\delta'}*e^{-\delta\frac{k+k'}{V}}*\\
*(e^{\delta\frac{k'}{V}}*(\delta\frac{k'}{V}-1)+1)*\frac{(1-\rho)\mu}{\delta+\mu(1-\rho)}
\end{multline}
\end{lemma}
\begin{IEEEproof}
Given the definition of $\theta|case2A$ seen in Equation \ref{eqn:theta2A}, its expected value can be expressed as:

\begin{multline}
E[\theta| case2A]=\frac{k'}{V}+E\left[\sum_{n=1}^{N_{2A}}T_{IC}(n)\right]+E[T_{IC}(0)]=\\
=\frac{k'}{V}+\sum_{n=0}^{\infty}\left(\sum_{i=1}^{n}E[T_{IC}(i)]*P\{N_{2A}=n\}\right)+E[T_{IC}(0)]
\end{multline}
From which, we can substitute the value obtained in Lemma \ref{lemN2A} to obtain:
\begin{multline}
E[\theta| case2A]=\\
=\frac{k'}{V}+\sum_{n=0}^{\infty}(\frac{n}{\delta'}*e^{-\delta\frac{k+k'}{V}}*\frac{(\delta\frac{k'}{V})^{n+1}}{n+1!}*\frac{(1-\rho)\mu}{\delta+\mu(1-\rho)} )+\frac{1}{\delta'}=\\
=\frac{k'}{V}+\frac{1}{\delta'}+\frac{1}{\delta'}*e^{-\delta\frac{k+k'}{V}}*\frac{(1-\rho)\mu}{\delta+\mu(1-\rho)}*\sum_{n=0}^{\infty}(\frac{n(\delta\frac{k'}{V})^{n+1}}{n+1!})
\end{multline}
With the series $\sum_{n=0}^{\infty}\frac{n*c^n}{n+1!}$ having result $e^c*(c-1)+1$, we obtain the formula of the lemma.
\end{IEEEproof}

\begin{lemma}
The expected value of $\theta| case2B$ is equal to: 
\label{AlemcaseB}
\begin{equation}
\label{AecaseB}
E[\theta| case2B]=\frac{k'}{V}\left(1+\frac{\delta}{\delta'}\right)
\end{equation}
\end{lemma}
\begin{IEEEproof}
Given the formulation of $\theta| case2B$ in Equation \ref{eqn:eqnt2b}, its expected value is found as:
$$E[\theta| case2B]=\frac{k'}{V}+E\left[\sum_{i=1}^{N_{2B}}T_{IC}(i)\right]$$
Similarly to Lemma \ref{lemcaseA}, we can find:
\begin{multline*}
E\left[\sum_{i=1}^{N_{2B}}T_{IC}(i)\right]=\sum_{n=0}^{\infty}\left(\sum_{i=1}^{n}E[T_{IC}(i)]\right) * P\{N_{IC}=n\}
\end{multline*}
that is equal, using Lemma \ref{AlemN}, to:
$$E\left[\sum_{i=1}^{N_{2B}}T_{IC}(i)\right]=\sum_{n=0}^{\infty}\frac{n}{\delta'} * e^{-\delta \frac{k}{V}}*\frac{(\frac{\delta k}{V})^n}{n!}=$$
$$=\frac{e^{-\delta \frac{k}{V}}}{\delta'}*\sum_{n=0}^{\infty}\frac{n*(\frac{\delta k}{V})^n}{n!}$$
Given that $\sum_{n=0}^{\infty}n*c^n/n!$ is a notable series with value $c*e^c$, we can write:
$$E\left[\sum_{i=1}^{N_{2B}}T_{IC}(i)\right]=\frac{e^{-\delta \frac{k}{V}}}{\delta'}*\frac{\delta k}{V}*e^\frac{\delta k}{V}=\frac{ k}{V}*\frac{\delta}{\delta'}$$
and returning to the starting formula, we have:
$$\theta| case2B=\frac{k'}{V}+\frac{k}{V}*\frac{\delta}{\delta'}$$
that is equal to the formula of the lemma.
\end{IEEEproof}

\begin{lemma}
\label{AlemcaseC}
The expected value of $\theta| case2C$ is equal to: 
\begin{multline}
\label{AecaseC}
E[\theta| case2C]= \frac{k'}{V}+E[T_{IC}(0)]+E\left[\sum_{i=1}^{N_{2C}}T_{IC}(i)\right]=\\
=\frac{1}{\delta'}+\frac{k'}{V}\left(1+\frac{\delta}{\delta'}\right)
\end{multline}
\end{lemma}
\begin{IEEEproof}
This result follows the line of reasoning of Lemma~\ref{AlemcaseB} with the addition of $E[T_{IC}(0)]$ (due to the fact that intercontact times are exponential, and therefore $E[T_{ICR}(0)] =
E[T_{IC}(0)]$), that gives us the formula in the lemma.
\end{IEEEproof}

\subsection{Analysis of data transfer in case 3}
In the third and final case (\emph{case 3}), we capture the scenario where the contact cannot last long enough to complete the input data transfer, that may happen for large inputs to transfer or short contact periods. In this case $B$ still starts during a contact periods, but, before the end of the time needed to transfer the input data ($k/V$), a first inter-contact period (denoted as $T_{IC}(0)$) happens. Then, after resuming the transmission, an additional number $RN_{IC} \geq 0$ of inter-contact periods $T_{IC}(i)$ may occur.  Therefore the formulation for $B | case3$ can be written as:
\begin{equation}
\label{eqn:equpcase3}
B | case3=\frac{k}{V}+T_{IC}(0)+\sum_{i=1}^{RN_{IC}}T_{IC}(i)
\end{equation}
To find the expected value of $B | case3$, we need a formulation for $P\{RN_{IC}=n\}$. This is not the same to the one seen for $P\{N_{2A}=n\}$ in Lemma \ref{lemN2A}, given that the elapsed time of the first contact period before the start of $B$ is unknown. We can then approximate the residual of the first contact with the entire contact duration, obtaining the following lemma.

\begin{lemma}
	\label{AlemRN}
	If there is at least a connection interruption during the input data transfer phase, the probability of having exactly $n$ additional interruptions, other than the first one, during the phase is equal to:
	\begin{equation}
	\label{AeRN}
	P\{RN_{IC}=n\}=\frac{ e^{-\delta \frac{k}{V}} \left(\delta \frac{k}{V}\right)^{n+1} }{n+1!}
	\end{equation}
\end{lemma}
\begin{IEEEproof}
	For the formulation of $P\{RN_{IC}=n\}$,  we consider it as the probability of $n$ contact periods $ \sum_{i=1}^{n} T_{C}(i)$ to be long enough to transfer the data that could not be transfered during the first contact period $\frac{k}{V} - T_{C}(0)$. Therefore $P\{RN_{IC}=n\}$ can be formulated as:
	\begin{multline}
	P\{RN_{IC}=n\}=\\
	=P\{ \sum_{i=1}^{n} T_{C}(i) < \frac{k}{V} - T_{C}(0) < \sum_{i=1}^{n+1} T_{C}(i) | T_{C}(0)<\frac{k}{V} \}
	\end{multline}
	Where, the summations can be substituted by the Erlang random variable $S_{C,n}$, obtaining:
	\begin{multline*}
	P\{RN_{IC}=n\}=\\
	= P\{S_{C,n} < \frac{k}{V} - T_{C}(0)< S_{C,n}+  T_{C}(n+1)|  T_{C}(0)<\frac{k}{V} \}
	\end{multline*}
	Again, we rewrite the expression in order to use the known formulas for the density function of Erlang distributions Equation \ref{eqn:erlang} and the cumulative probability function for exponential distributions Equation \ref{eqn:cumulc}:
	\begin{multline*}
	P\{RN_{IC}=n\}=\\
	=\int_0^\frac{k}{V} P\{S_{C,n} < \frac{k}{V} - t < S_{C,n}+  T_{C}(n+1)\} * f_{T_{C}(0)}(t) dt =
	\end{multline*}
	$$=\int_0^\frac{k}{V} \int_0^{\frac{k}{V}-t}P\{\frac{k}{V} - t < x+ T_{C}(n+1) | S_{C,n}=x \} *$$
	$$* f_{S_{C,n}}(x) dx * f_{T_{C}(0)}(t) dt =$$
	$$=\int_0^\frac{k}{V}P\{\frac{k}{V} - x < T_{C}(n+1) | S_{C,n}=x\}*f_{S_{C,n}}(x) dx= $$
	$$=\int_0^\frac{k}{V} \int_0^{\frac{k}{V}-t} (1-F_{T_{C}(n+1)}(\frac{k}{V}-t-x))*$$
	$$*f_{S_{C,n}}(x)*f_{T_{C}(0)}(t) dx dt=$$
	Substituting the values, we obtain:
	\begin{multline*}
	P\{RN_{IC}=n\}=\int_0^\frac{k}{V} \int_0^{\frac{k}{V}-t} e^{-\delta(\frac{k}{V}-t-x)}*\\
	*\frac{\delta^n x^{n-1} e^{-\delta x}}{n-1!}*\delta e^{-\delta t}dx dt=\\
	\end{multline*}
	$$= \frac{ e^{-\delta\frac{k}{V}} \delta^{n+1} }{n-1!} * \int_0^\frac{k}{V} \int_0^{\frac{k}{V}-t} x^{n-1} dx dt=$$
	$$=  \frac{ e^{-\delta\frac{k}{V}} \delta^{n+1} }{n-1!}  \int_0^\frac{k}{V} \frac{(\frac{k}{V}-t)^n}{n} dt=$$
	That, through integration, is equal to:
	$$=\frac{ e^{-\delta\frac{k}{V}} (\delta \frac{k}{V})^{n+1} }{n+1!}$$
\end{IEEEproof}

Thanks to Lemma \ref{AlemRN}, we can find the expected value of $B | case3$:
\begin{lemma}
	\label{Alemcase3}
	The expected value of a data transfer phase that starts during a contact period, but having at least one disconnection period, is equal to:
	\begin{equation}
	E[B | case3]=\frac{k}{V}\left(1+\frac{\delta}{\delta'}\right)+\frac{e^{-\frac{\delta k}{V}}}{\delta'}
	\end{equation}
\end{lemma}

\begin{IEEEproof}
	From the definition of $B | case3$, its expected value is calculated as:
	$$E[B | case3]=\frac{k}{V}+E\left[\sum_{i=1}^{RN_{IC}+1}T_{IC}(i)\right]=$$
	\begin{equation}
	= \frac{k}{V}+\sum_{n=0}^{\infty}\sum_{i=1}^{n+1}E[T_{IC}(i)] * P\{RN_{IC}=n\}
	\end{equation}
	Substituting the value of $P\{RN_{IC}=n\}$ provided by Lemma \ref{AlemRN}, we obtain:
	$$E[B | case3]=\frac{k}{V}+\frac{1}{\delta'}+\sum_{n=0}^{\infty}\frac{n}{\delta'}*\frac{ e^{-\delta\frac{k}{V}} (\delta \frac{k}{V})^{n+1} }{n+1!}=$$
	$$=\frac{k}{V}+\frac{1}{\delta'}+\frac{ e^{-\delta\frac{k}{V}} }{\delta'}*\sum_{n=0}^{\infty}\frac{n (\delta \frac{k}{V})^{n+1}}{n+1!}$$
	we then divide the series in two notable exponential series:
	\begin{multline*}
	E[B | case3]=\\
	=\frac{k}{V}+\frac{1}{\delta'}+ \frac{e^{-\delta\frac{k}{V}}}{\delta'}*\sum_{n=0}^{\infty}\frac{(n+1) (\delta \frac{k}{V})^{n+1}}{(n+1)!}-\sum_{n=0}^{\infty}\frac{(\delta \frac{k}{V})^{n+1} }{(n+1)!}=\\
	=\frac{k}{V}+\frac{1}{\delta'}+\frac{e^{-\delta\frac{k}{V}}}{\delta'}  * (\delta\frac{k}{V}e^{\delta\frac{k}{V}}-(e^{\delta\frac{k}{V}}-1))=\\
	=\frac{k}{V}(1+\frac{\delta}{\delta'})+e^{-\delta\frac{k}{V}}*\frac{1}{\delta'}\\
	\end{multline*}
\end{IEEEproof}

To formulate $\theta |case3$, likewise to what we have seen in cases 2B and 2C, we divide the formulation into in two sub-cases 3A and 3B depending on the connection state between the seeker and the provider at the start of the phase.

In case3A the output data transfer phase starts during a contact and in case3B during an inter-contact period, without any assumption on the number of disconnections occurring in the phase. So, $\theta |case3A$ can be formalized exactly as seen in Equation \ref{eqn:eqnt2b} and $\theta |case3B$  can be formulated as in Equation \ref{eqn:eqnt2c}.

For simplicity, in this case we don't keep track in the analysis of the time evolution of the previous phases of service provisioning time with respect to the contact and inter-contact processes. Therefore, we approximate $p_{case3A}$ and $p_{case3B}$ using the steady state probabilities of contact and inter-contact phases:
\begin{equation}
p_{case3A}=\frac{E[T_{C}]}{E[T_{C}]+E[T_{IC}]}
\end{equation}
\begin{equation}
p_{case3B}=\frac{E[T_{IC}]}{E[T_{C}]+E[T_{IC}]}
\end{equation}
With the formulation of $\theta$ in the two sub-cases and their probability, we obtain the following lemma:

\begin{lemma}
	The expected value of $\theta|case3$ is 
	\begin{equation}
	\label{eqn:ethetacase3}
	E[\theta|case3]=\frac{k'}{V}\left(1+\frac{\delta}{\delta'}\right)+\frac{\delta}{\delta'(\delta'+\delta)}
	\end{equation}
\end{lemma}
\begin{IEEEproof}
	This immediately follows by using the results in Lemma \ref{AlemcaseB} and Lemma \ref{AlemcaseC} to then apply the law of total probability.
\end{IEEEproof}

\subsection{Derivation of $p_1$}
\begin{lemma}
	\label{Alemp1}
	The probability $p_1=P\{R<T_{C}(0)\}$ of a single-service request resolution process, involving a seeker $j$, a provider $h$ and the execution of service $s_i$, ending during the same contact event when the input data transfer phase started can be approximated as:
	\begin{equation}
	p_1=\frac{\mu (1-\rho)e^{-\delta \frac{k+k'}{V}}}{\delta + \mu (1-\rho)}
	\end{equation}
\end{lemma}
\begin{IEEEproof}
	This is straightforward by recalling that, in case 1, $R=\frac{k}{V}+DS_{s_i,j}+DQ_{j}+\frac{k'}{V}$.  We can condition on the value of $DS_{s_i,j}+DQ_{j}$ and apply the law of total probability. The distribution of $DS_{s_i,j}+DQ_{j}$ is derived in Equation \ref{eqn:Aeserv} in the Appendix. We obtain:
	\begin{multline}
	P\{R<T_{C}(0)\}=\\
	=P\left\{\frac{k}{V}+DS_{s_i,j}+DQ_{j}+\frac{k'}{V} < T_{C}(0)\right\}=\\
	= \int_{0}^{\infty}\left(1-F_{T_{C}}\left(\frac{k+k'}{V}+t\right)\right)*f_{DS_{s_i,j}+DQ_{j}}(t)dt$$
	\end{multline}
	that gives the formulation of $p_1$.
\end{IEEEproof}

\subsection{Further analysis of the model accuracy and MEV components}
\label{sub:validation}
In a simulated opportunistic computing environment (as described next)
we use the algorithm proposed in this paper to select compositions, and
compare the estimates of the model for minimum service composition with the
values observed in simulations for the selected compositions.
%\cite{pmtr-2008}

\begin{table}[t]
	\caption {Synthetic-trace simulation parameters}
	\centering{
		\begin{tabular}{|p{4.5cm}|p{3cm}|}
			\hline
			Simulation runs & $5$\\
			\hline
			Number of nodes & $30$\\
			\hline
			Simulation space & $500m \times 500m$\\
			\hline
			Total simulation time & $400000s$\\
			\hline
			Mobility warm-up period & $10000s$\\
			\hline
			Connectivity range & $90m$\\
			\hline
			Transmission speed & $2Mbps$\\ 
			\hline
			RWP node speed & $[0.6 , 2.26]m/s$\\
			\hline
			RWP node pauses & $0s$\\
			\hline
			Density of each service & $25\%$\\
			\hline
			Input type range & $i \in [0,7]$\\
			\hline
			Requests output type range & $o \in [1,8], o>i$\\
			\hline
		\end{tabular}
	}
	\label{tab:SimPar}
	\vspace{-12pt}
\end{table}

To validate more precisely the different stages of the model and explore a
larger range of parameters, we used synthetic mobility traces. Specifically, we
used the RandomWayPoint model, modified as discussed in \cite{Navidi2004} in
order to avoid problems related to the initial transient phase of the mobility
process\footnote{Note that, although in
	general other mobility models are considered more realistic, RWP is still a
	valid option when users form a
	unique social community moving in a common area \cite{Boldrini2010}.}.
The main simulation
parameters are described in Table~\ref{tab:SimPar}. The total simulation time is
400000s. During the first 10000s nodes only
collect information about
their average contact and inter-contact times. 

\begin{figure}[tb]
	\centering
	\includegraphics[width=0.45\textwidth,height=0.25\textwidth]{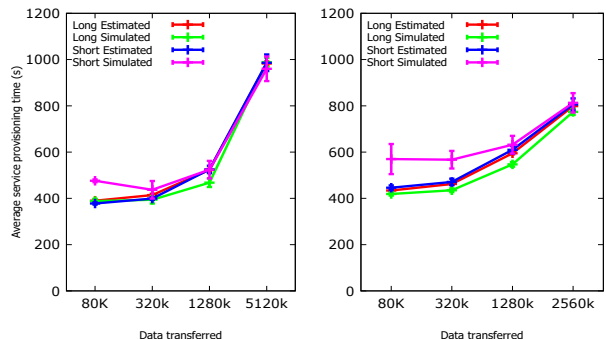}
	\caption{Service provisioning time (short/long warmup)}
	\label{fig:comp1}
	\vspace{-12pt}
\end{figure}

Such relatively long warm-up periods are useful to test the performance of the proposed algorithm when ample time is allowed for collecting statistics across nodes. To test its sensitiveness to shorter warm-up periods we have also run simulations where the simulation is 28800s (8 hours, a typical working day) long, and the warm-up is only 1000s. In both cases, we used two different rate of request generation, labelled hereafter "5-8" and "20-40", where the time between two consecutive service requests follows a uniform distribution in the range [5,8]s or [20,40]s respectively. In these experiments, Input/Output parameter sizes vary from 80 KBytes up to 2560 KBytes for the "5-8" case and from 80 KBytes up to 5120 KBytes for the "20-40" case\footnote{This difference is due to the occurrences of overflowing in the nodes' output buffers caused by the limitations of the simulator when high amount of data is transferred on a high load scenario.}. We decided to present these values for the transfer sizes, as in the considered simulated scenarios they generate a light and high traffic load during encounter between nodes, and are thus representative of the two ends of the spectrum, i.e. a non-saturated and a saturated network condition. Figure \ref{fig:comp1} shows the average service provisioning time for the different request rates, and for increasing sizes of input/output parameters. Curves obtained with long and short warm-up periods are labelled as "Long" and "Short", respectively.

We can see that, in both cases, the simulated service provisioning times are lower for the "Long" cases, with a clear difference for small input and output data sizes. In the "Long" case estimates differ at most by 11\% from the simulated times for light request load and at most by 8\% for heavy load, while for the ``Short" case estimates are up to 20\% higher than in simulations. While there is a clear difference, we can anyway conclude that even relatively short warm-up periods are sufficient to obtain reasonable estimates. The following results are obtained using long warm-up periods.

To validate the components of the model, we consider two cases where we use, respectively,  only single-component services (named ``Single'') and only composed services (named ``Comp'') because single-component services are not available in the network. 
For both scenarios we run two sets of simulations for the ``5-8'' and ``20-40'' settings respectively.

\begin{figure}[tb]
	\centering
	\includegraphics[width=0.45\textwidth,height=0.25\textwidth]{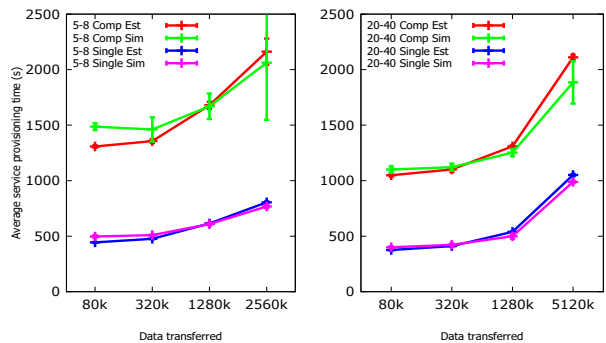}
	\caption{Service provisioning time (single/multi component)}
	\label{fig:comp2}
	\vspace{-18pt}
\end{figure}

In Fig. \ref{fig:comp2}, we can see the average service provisioning
time in simulation (\emph{Comp Sim} and
\emph{Single Sim}) and the estimates provided by the model (\emph{Comp Est} and \emph{Single Est}) for the ``5-8'' and ``20-40'' scenarios.
For the ``5-8'' scenario the model is very accurate in the ``Single'' case. When
compositions are used, simulation results have a much larger variance,
as expected, and are  more difficult to
precisely model in case of high request generation
rates, due to the higher variability of statistics about the status of
providers in case of higher loads. Nevertheless, it is clear that the
model accurately follows the trend of the
simulation results. The difference of service time between the
``Single'' and ``Comp'' cases is due to the fact that the amount of services executed in ``Comp'' is almost double
and that there is at least one more
data transfer between nodes for each request.

For the ``20-40'' scenario we see that with a lower generation rate, the network is in general less loaded.
The effect is particularly evident in the case of compositions. Average
completion times are reduced by about 500s, and  the maximum difference between
estimated and real values is 13\% for the case of compositions and 11\% for the
case of single service execution.

\begin{figure}[t]
	\centering
	\includegraphics[width=0.45\textwidth,height=0.25\textwidth]{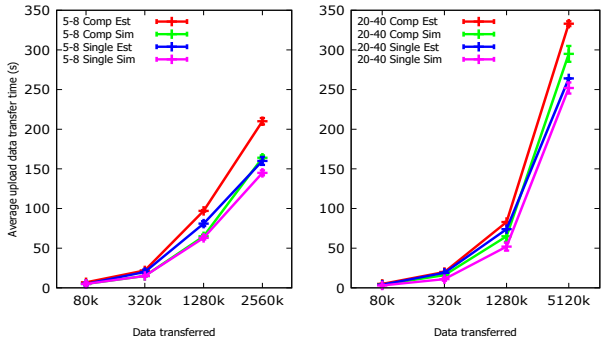}
	\caption{Average upload transfer time}
	\label{fig:upl}
	\vspace{-12pt}
\end{figure}
\begin{figure}[t]
	\centering
	\includegraphics[width=0.45\textwidth,height=0.25\textwidth]{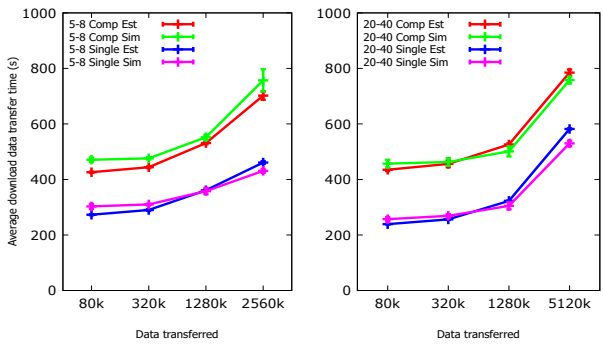}
	\caption{Average download transfer time}
	\label{fig:down}
	\vspace{-18pt}
\end{figure}

In Fig. \ref{fig:upl}  we show the results for the upload phases. We can see how
in the two cases we obtain good estimates. We observe a higher overestimate
between the model and the simulation results for the "Comp" case, in particular
when the load increases (scenario "5-8") and for higher sizes of the parameters.
In general, transfer times increase significantly with the size of the
parameters, which is expected. Note that as the size increases, it becomes more
and more likely that a single contact is not sufficient to complete a
transfer. This results in a higher probability that also inter-contact times
must be factored in the upload time. %Given inter-contact times are quite long

In Fig. \ref{fig:down} we show the results for the download phases. The first
aspect that can be noticed is that the model is quite accurate, with a maximum error in the order of 10\%. In this case, we do
not observe a very steep increase of the curves for large parameter sizes, while the download times are larger
than upload times for small parameter sizes. This is because the download phase
typically always includes at least an initial (residual) inter-contact time,
because the seeker and provider are most of the time not in contact when service
execution is over.

\end{document}